\documentclass[aps,preprint,nofootinbib]{revtex4}
\usepackage{graphicx}
\usepackage{color}
\usepackage{epsfig}
\usepackage{subfig}
\usepackage{caption}
\usepackage[symbol]{footmisc}

\begin{document}
\begin{center}
	\textbf{Liquid-Liquid Phase Transition in Supercooled Silicon\footnote[2]{Review article published as a chapter in "Liquid Polymorphism", The Advances in Chemical Physics, Vol 152, 463-517, 2013 (Wiley).}}\\
	\textit{Vishwas V Vasisht$^{1}$ and Srikanth Sastry$^{1,2}$\\}
	\textit{$^{1}$ Theoretical Sciences Unit, Jawaharlal Nehru
          Centre for Advanced Scientific Research, Jakkur Campus,
          Bangalore 560 064, India.}
	\textit{$^{2}$ TIFR Centre for Interdisciplinary Sciences, Tata Institute of Fundamental Research, 21 Brundavan Colony, Narsingi, Hyderabad 500 075, India.\\}
\end{center}
        
\bigskip

\tableofcontents
\begin{center}
\line(1,0){300}
\end{center}

\section{Introduction}
Silicon, the second most abundant element in the earth's crust, is ubiquitous in the form
of silica and silicates in the natural world. In the elemental form, it is an essential
component of the semiconductor technology. It was first prepared in its amorphous form by
J.J. Berzelius and later the crystalline form by H.E. Sainte-Claire Deville
\cite{Weeks_JCE_1932} in the 1800's. The crystalline and amorphous solid are the two most
familiar forms of silicon, which have been studied extensively. The crystalline form of
silicon is a tetravalent semiconductor (as is the amorphous solid) and upon melting at
$1687 K$ at ambient pressure, transforms to a metallic liquid with higher coordination
number, around 6. Liquid silicon is relatively less studied, given the elevated
temperatures at which it exists. Nevertheless, it has been a subject of substantial
experimental, theoretical and computational investigation, both at temperatures above the
melting temperature, and in the supercooled and stretched (negative pressure) states. The
investigations of the metastable liquid have been motivated, as this review seeks to
demonstrate, by fundamental questions regarding (i) the eventual fate of metastable
liquids upon deep undercooling and stretching, (ii) the interest in the possibility of a
novel transition between two distinct liquid forms in a class of ``tetrahedral'' liquids
to which silicon belongs, and (iii) the role of the thermodynamics of metastable liquid
states on the kinetics of phase transformations, particularly to the crystalline state.

Based on the extrapolated Gibbs free energies of amorphous solid and liquid phases for
germanium, and a scaling of temperatures for the case of silicon, Bagley and Chen
\cite{bagley_AIP_1979} and independently Spaepen and Turnbull \cite{spaepen_AIP_1979}
suggested a first order phase change from the amorphous solid to the liquid near $T=1349 K$
for silicon, below the freezing point of liquid $T_m=1687 K$. Subsequent experimental as well
as simulations studies \cite{donovan_APL_1983, Thompson_PRL_1984, Broughton_PRB_1987,
Luedtke_PRB_1989, Cook_PRB_1993, Wautelet_PSSB_1990, Stich_PRB_1991, Rosato_JAP_1999,
Rosato_CMS_2000, Morishita_PRL_2004} of non-crystalline silicon supported this notion
though the precise nature of the transition was unclear owing to the limited availability of
information. Using a two-state model, Aptekar \cite{aptekar_sovphydok_1979,
Ponyatovsky_MSR_1992} in 1979 proposed a phase diagram that described the liquid and
amorphous states as {\it two states of one noncrystalline (liquid) phase}, and further
predicted a negative pressure critical point. The perspective that the transition from
amorphous silicon to the liquid should be viewed as a liquid-liquid, rather than an
(amorphous)solid-liquid transition, attracted renewed interest in light of an independent proposal of
a liquid-liquid transition in the case of water \cite{poole_nature_1992}, and the growing
appreciation that such a possibility was in principle also applicable to other substances
such as, {\it e. g.} silica \cite{Ivan_PRE_2000}, which exhibited thermodynamic and
structural features similar to water.

The analogy was pursued by Angell and co-workers \cite{angell_JNCP_1996} who proposed a
first order liquid-liquid transition line as a feature in the pressure-temperature phase
diagram of silicon, based on simulation evidence using the Stillinger-Weber ($SW$) potential
of silicon \cite{stillinger_weber_PRB_1985}.  The experimental work of Deb and co-workers
\cite{Deb_nat_2001} observed a pressure induced amorphous-amorphous transition and
speculated a possibility of an underlying liquid-liquid transition ($LLT$), to be found at
ambient pressure at around ($1400 K$). From extensive simulations of silicon using the SW
model potential Sastry and Angell \cite{sastry_nmat_2003} found evidence of a
liquid-liquid transition at zero pressure at around $1060 K$, which was also subsequently
supported by {\it ab initio} simulations \cite{Jakse_JCP_2008, Ganesh_PRL_2009}. Considerable
recent simulation and experimental evidence lends support to the idea of a liquid-liquid
phase transition at ambient pressures in silicon \cite{Deb_nat_2001, sastry_nmat_2003,
Miranda_JCP_2004, hedler_nature_2004, ashwin_prl_2004, McMillan_nmat_2004,
McMillan_nmat_2005, Jakse_PRL_2007, Jakse_JCP_2008, Ganesh_PRL_2009, Beye_PNAS_2010,
Sastry_PNAS_2010, Vasisht_nphys_2011, Garcez_JCP_2011, sciortino_nphy_2011,
Ganesh_JNCS_2011, Okada_PRL_2012}. However, such a transition is also consistent with a ``critical point
free'' scenario, proposed recently by Angell \cite{angell_Science_2008}, and hence, the
existence of a critical point needs independent verification. The simulation evidence of a
critical point at negative pressures has been provided by Vasisht {\it et al.}
\cite{Vasisht_nphys_2011, Vasisht_nphys_sup_2011} for SW silicon and in {\it ab initio}
simulations by Ganesh and Widom \cite{Ganesh_PRL_2009}. The work of Vasisht {\it et al.}
also maps out the detailed phase diagram of supercooled silicon, finding phase behaviour
that is similar in qualitative features to that found in similar studies of water and
silica. Changes in the electronic structure accompanying this phase transition have also been studied
computationally \cite{ashwin_prl_2004, Jakse_PRL_2007, Ganesh_PRL_2009}, demonstrating a transition from a
metallic high temperature liquid to a low temperature liquid with substantially reduced conductivity.

The introductory outline above spans investigations pursued over more than three
decades and is inevitably not comprehensive. In the following sections, we elaborate on
some of these themes. While we present salient experimental, theoretical, and {\it ab
initio} simulation results, the main body of results discussed will be from classical
computer simulations of SW silicon. We do not make an attempt to be exhaustive, and
regret the omission of any significant material concerning the topic of this review. We
broadly divide the investigations described into those that address an ``amorphous solid
to liquid'' transition, and those that address a ``liquid-liquid'' transition in
supercooled silicon. Correspondingly, the next section describes early work that largely
falls into the first category, although including the work of Aptekar
\cite{aptekar_sovphydok_1979} and Ponyatovsky and Barkolov \cite{Ponyatovsky_MSR_1992}
which address a liquid-liquid transition. The parallel developments in the case of water
(and later by extension, other tetrahedral liquids) were aimed at developing an
explanatory framework for its anomalous properties. Therefore in Section III, we describe
the various scenarios put forward for fluids exhibiting density and related anomalies. In
Section IV, we discuss the more recent investigations, starting with the work of Angell
and co-workers \cite{angell_JNCP_1996} which explore the similarities in the phase
behaviour of silicon with water and other analogous substances. We discuss in this section
the details of the phase diagram as revealed by simulations of SW silicon including a
negative pressure critical point \cite{Vasisht_nphys_2011} as well as structural and
dynamical properties of supercooled silicon. An important feature of the
behaviour of silicon, not necessarily shared with other substances that may exhibit a
liquid-liquid transition is the change in electronic structure, from a semiconductor at
low temperatures to a metal in the high temperature liquid state. We therefore discuss the
electronic structure in Section V separately, including electronic structure calculations
performed for structures obtained in classical simulations \cite{ashwin_prl_2004}, first
principle molecular dynamics ($FPMD$) simulation studies \cite{Jakse_JCP_2008,
Ganesh_PRL_2009} and experimental work of Beye and co-workers \cite{Beye_PNAS_2010} which
employ the changes in the electronic density of states as an experimental probe for the
liquid-liquid transition in silicon. Since a significant part of the simulation results
presented are based on a classical empirical potential (performed to a large extent by the
authors of this review), an important question to address is the reliability of this
potential in describing the behaviour of silicon. To this end, in Section VI we compare the
structural, dynamical and thermodynamic quantities obtained from the SW potential with those from
{\it ab initio} simulations and with available experimental data, to provide a critical
assessment of the applicability of classical simulation results to real silicon.  We also
discuss the sensitivity of the thermodynamic properties to model parameters. We end with a
summary in Section VII.

\section{Early Work on Metastable Silicon}

\begin{figure}[t]
	\begin{center}
		\includegraphics[scale=0.25]{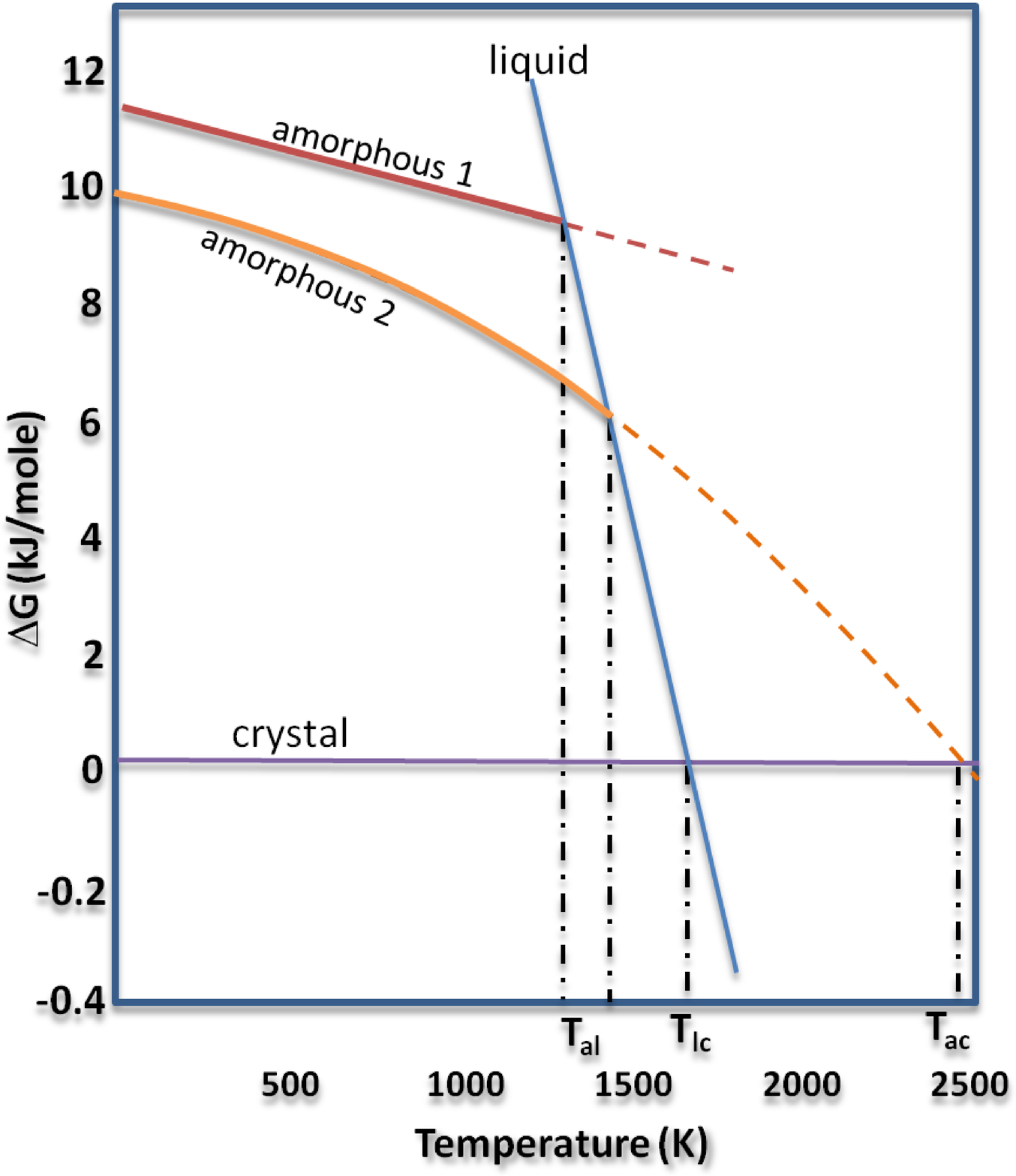}
		\caption{The estimate of excess Gibbs free energy ($\Delta G$) against temperature
		suggesting a first order transition.  The brown and orange solids lines represent
		two extreme estimates of $\Delta G$ for the amorphous phase and the dashed lines are
		extrapolation of these $\Delta G$ into the liquid phase. The blue line shows $\Delta
		G$ for the liquid phase and the purple line is the reference crystal phase value.
		$T_{al}$, $T_{lc}$ and $T_{ac}$ represent liquid-amorphous, crystal-liquid and
		crystal-amorphous phase transition temperatures respectively. [Adapted from Donovan
		{\it et al.} \cite{donovan_JAP_1985} with permission.]}
		\label{fig:Donovan_FE}
	\end{center}
\end{figure}

The earliest experimental work on metastable silicon includes studies carried out by
Bagley and Chen \cite{bagley_AIP_1979}, Spaepan and Turbull \cite{spaepen_AIP_1979},
Donovan {\it et al.} \cite{donovan_APL_1983} and Thompson {\it et al.}
\cite{Thompson_PRL_1984}. Bagley and Chen \cite{bagley_AIP_1979}, and independently
Spaepen and Turnbull \cite{spaepen_AIP_1979}, used the available thermodynamic data for
germanium (such as the heat capacities measured by Chen and Turnbull
\cite{Chen_JAP_1969}), and the kinetics of crystal growth, to estimate the temperature
dependence of the excess Gibbs free energy of amorphous and liquid silicon. The
significantly different entropies of the two limiting states leads to the prediction (by
extrapolation) of a slope discontinuity in the Gibbs free energy, or a first order phase
transition from a four coordinated amorphous to a metallic liquid state. The estimated
transition temperature was $T_{al}$ at $1349 K$ for silicon (by scaling the melting
points of germanium and silicon). Later Donovan and co-workers
\cite{donovan_APL_1983,donovan_JAP_1985} performed differential scanning calorimeter
($DSC$) measurements on amorphous silicon, produced by ion implantation
and based on Gibbs free energies deduced, estimated $T_{al}$ to be $1420K$. We shown in
FIG. \ref{fig:Donovan_FE} the Gibbs free energies estimated by Donovan {\it et al.}
\cite{donovan_JAP_1985}. This transition was also confirmed by another experimental group
of Thompson and co-workers \cite{Thompson_PRL_1984} who performing pulsed-laser melting of
amorphous silicon and estimated the $T_{al}$ as $1480 \pm 50 K$.


The theoretical analysis of the thermodynamics of supercooled silicon, presented by
Aptekar \cite{aptekar_sovphydok_1979} treats the liquid as a pseudo-binary regular
solution of two components, along lines explored in related contexts by Rappaport
\cite{Rapoport_1967_JCP}, Ponyatovsky and co-workers \cite{Ponyatovsky_MSR_1992,
Ponyatovsky_JPCM_2003}. The two components are characterised by different local bonding
environments (covalent or metallic). Correspondingly, the Gibbs free energy of the liquid
is written as

\begin{equation} 
G_{l} = G_{1} (1 - \omega) + G_{2} \omega + W (1 - \omega) \omega + R T [\omega  \log \omega  + (1 - \omega ) \log (1 - \omega )].
\end{equation} 

The parameter $\omega$ describes the {\it degree of metallisation} and is determined by
the equilibrium condition of the liquid. Writing the free energy difference between the
two pure liquids $G_{1}$ and $G_{2}$ phenomenologically, and using values for the various
parameters involved from available experimental data, Aptekar estimated phase diagrams for
germanium and silicon, showing in each case that the liquids exhibit metastable, negative
pressure, liquid-liquid critical points. Although this analysis is motivated by high
pressure transformation of semiconductors to metallic states, this analysis also offers
a rationalisation of results concerning the transformation of amorphous silicon to liquid
upon heating.



\begin{figure}[b]
	\begin{center}
		\includegraphics[scale=0.5]{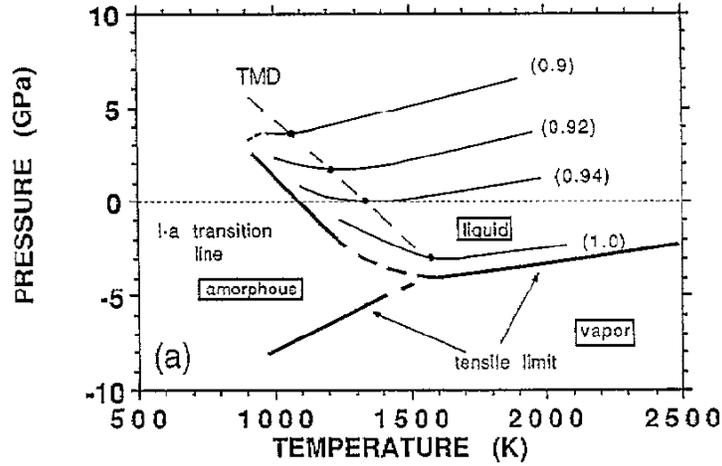}
		\caption{Phase diagram proposed by Angell and co-workers \cite{angell_JNCP_1996}
		based on simulations of the SW potential, with a liquid-amorphous transition line
		that is negatively sloped. Also shown are the locus of density maxima and the
		tensile limit line. [From Angell {\it et al.}  \cite{angell_JNCP_1996} with
		permission.]}
		\label{fig:Angell_SW_PD}
	\end{center}
\end{figure}

Given the difficulties in experimental studies of extreme states of matter (high
undercooling, high temperatures, pressures {\it etc}), computer simulations have, over the
last few decades, helped gain insights into states that are hard to probe experimentally.
Unlike in experiments, it is a relatively simpler task to explore a wide range of
temperatures and pressures in simulations and thus bracket the region of interest for
further experimental verification. For more than three decades computer simulation studies
of silicon have been carried out using various empirical interaction potentials (a
comprehensive study of six different empirical potential is given in the reference
\cite{balamane_PRB_1992}) and also using first principle simulations \cite{Stich_1989_PRL,
Morishita_PRL_2006, Jakse_PRL_2007, Ganesh_PRL_2009}. One of most widely used potential
\cite{Broughton_PRB_1987, Luedtke_PRB_1989, angell_JNCP_1996, sastry_nmat_2003,
ashwin_prl_2004} for studying silicon in computer simulations is the Stillinger-Weber
($SW$) potential \cite{stillinger_weber_PRB_1985}. Using the SW potential in molecular
dynamics ($MD$) simulations, Broughton and Li \cite{Broughton_PRB_1987} performed one of the
earliest studies of the liquid, crystal and amorphous phase diagram of silicon. In this
work Broughton and Li found that the crystal and liquid phases are well represented by the
SW potential, but the thermodynamics of the amorphous phase is poorly described and that
the supercooled liquid phase does do not undergo a first order transition to an
amorphous state upon cooling. Luedtke and Landman \cite{Luedtke_PRB_1989} showed that this
failure to obtain amorphous Si {\it via} direct cooling of the melt in simulations is
related to the quench rates employed. These authors \cite{Luedtke_PRB_1989} noted that upon cooling, a sharp change
in the energy and density of the system occurred at $T\sim1060 K$ followed by a slow
variation in these properties as cooling continues to $T=300 K$. These authors
compared their system obtained from quenching with the amorphous phase obtained from an
alternate method (involving tuning of coefficient of three body part of the SW potential)
and found that both the systems have comparable structural composition. Angell and
co-workers \cite{angell_JNCP_1996} using the SW potential explored a relatively wide range
of temperature and pressure to chart out the phase diagram of metastable silicon. These
authors also reported a well defined transition, at $T \sim 1060 K$, from a highly diffusive liquid
states to a non-diffusive ``amorphous phase'', with a coordination number of $4.1$. The phase
diagram as suggested by Angell and co-workers is shown in the FIG. \ref{fig:Angell_SW_PD} (note that
at higher pressures, the transition occurs at lower temperatures, leading to a negatively sloped
transition line).

As noted earlier, the work of Angell and co-workers \cite{angell_JNCP_1996} made contact
with the possibility of a liquid-liquid transition in the case of water, which had been
proposed as one of the possible scenarios within which to understand the anomalous
properties of water. Since these scenarios are relevant for our further discussion, we
review them briefly in the next section before returning to more recent results for
silicon.



\section{Scenarios for Liquids Displaying Thermodynamic Anomalies}

It is well known that ice floats on water, owing to the solid form having a lower density
than the liquid. It is also well known that liquid water has the maximum density at
ambient pressure at $4^\circ C$, below which the density {\it decreases} as temperature
decreases, contrary to the normal behaviour wherein liquids become denser as the
temperature decreases. Hence, the decrease of density of water with temperature is
described as {\it anomalous}. The temperature at which the density is a maximum depends on
pressure, and thus once has a locus of temperatures of maximum density ($TMD$) which for
water at positive pressures is negatively sloped in the ($P, T$) plane.  Liquid water also
shows anomalous behaviour in thermodynamic quantities like compressibility $(K_T)$ and
heat capacity $(C_P)$. The compressibility of water decreases with a decrease in temperature
like any other liquid, but reaches a minimum at $46^\circ C$. Below this temperature the
compressibility increases with a decrease in temperature and shows an apparent divergence at
$-45^\circ C$ \cite{Speedy_JCP_1976} when only the anomalous component of the
compressibility is considered. The specific heat capacity of liquid water increases with a
decrease in temperature and passes through a minimum at $36^\circ C$ and shows normal
behaviour at higher temperatures. At around $-47^\circ C$ the specific heat displays an
apparent divergence \cite{Angell_JPC_1982}. Water also shows anomalous behaviour in its
dynamical properties. The diffusivity of liquid water increases with increase in pressure
which is abnormal for liquids. The work of Errington and Debenedetti
\cite{errington_nat_2001} identifies a region in the phase diagram where the structure of
the liquid behaves anomalously. This work also found that the anomalies in density and diffusivity 
occur within the structurally anomalous region (in the $\rho-T$ phase diagram of water).

In the case of silicon, even though simulations have predicted anomalies similar to that of
water, experimentally none of the anomalies have been verified. The density maximum in
silicon, at zero pressure, as predicted by classical SW simulations is at $T=1350 K$
\cite{angell_JNCP_1996, Sastry_PhyA_2002, Vasisht_nphys_2011} and {\it ab initio} simulations
predict it at around $T=1200 K$ \cite{Morishita_PRL_2006}. The lowest temperature at which density
measurements have been carried out in experiment is $T=1370 K$ \cite{rhim_JCG_2000}.  An
extrapolation of a polynomial fit to the experimental data from \cite{rhim_JCG_2000} would suggest a
density maximum around $1200 K$. In SW silicon simulations density minima have also been identified
\cite{Vasisht_nphys_2011}. A very recent experimental measurement of isobaric heat capacity
\cite{Kobatake_MST_2010} has been carried out down to $T=1548 K$ which shows an increase in heat
capacity with a decrease in temperature similar to that of water below $T=309 K$. The SW silicon
simulations predict a very weak minimum in $C_P$ at around $T=3350 K$. There are no experimental reports of
compressibility but in simulations of SW silicon, both compressibility maxima and minima have been
identified
\cite{Vasisht_nphys_2011}. Along with these thermodynamic anomalies, silicon is also found
to show anomalous behaviour of self diffusion. The diffusivity of supercooled liquid silicon is found
to increase with increase in compression \cite{Morishita_PRE_2005, Vasisht_nphys_2011},
which is yet to be verified in experiments. Some of these anomalies have also been seen in
the case of silica \cite{poole_PRL_1997, Ivan_PRE_2000, shell_PRE_2002} 


Various models and scenarios (based on thermodynamic constraints) have been developed to
explain the thermodynamic anomalies of water \cite{debenedetti_book_1996,
Mishima_nat_1998, Brovchenko_CPPM_2008}, (and by extension, other liquids with water-like
anomalous behaviour, including silicon) which are briefly discussed in this section. The
observation of negative melting curves in various systems including water and silicon,
means (from the Clausius-Clapeyron relation $dP/dT = \Delta S_m/\Delta V_m$, and assuming
that the entropy of the crystal is lower than that of the liquid) that the liquid density
will be greater than that of the solid phase. This is a feature that is typical of the
substances to which the considerations in this section apply.

Below, we describe some of the scenarios that have been explored as a way of
rationalising the thermodynamics of liquids displaying anomalies, such as water. These
are: (i) the stability limit conjecture \cite{Speedy_JCP_1976, Speedy_JPC_1982,
debenedetti_AIChE_1988, debenedetti_book_1996}, (ii) the liquid-liquid critical point
scenario \cite{poole_nature_1992},  (iii) the singularity free scenario
\cite{sastry_pre_1996} and (iv) the critical point free scenario
\cite{angell_Science_2008}.

\begin{figure}[t]
	\begin{center}
		\includegraphics[scale=0.15]{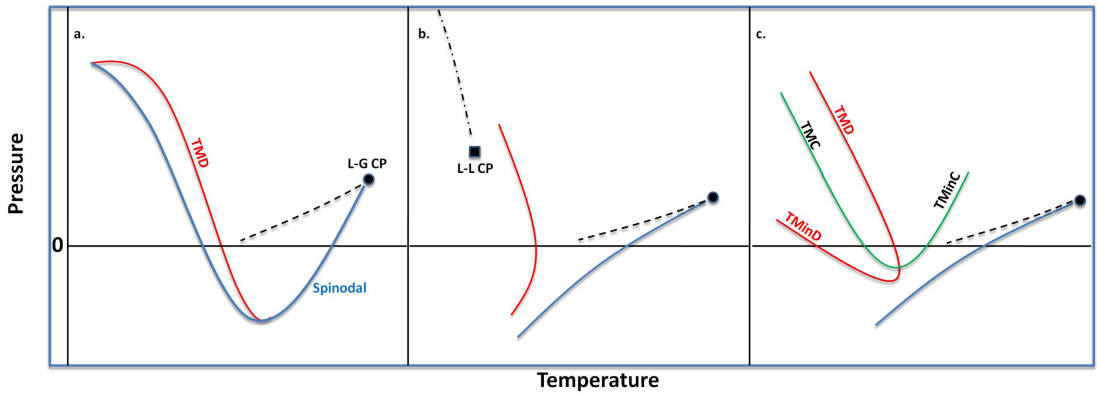}
		\caption{Schematic phase diagrams in the pressure-temperature (P, T) plane
		illustrating three scenarios for liquids displaying anomalous thermodynamic
		behavior. (a.)  The spinodal retracing scenario. (b.) The liquid-liquid critical
		point scenario. (c.) The singularity free scenario. The green lines represent the
		locus of compressibility extrema, and the red lines the locus of density extrema.
		The dashed and dot-dashed lines represent liquid-gas and liquid-liquid transition
		lines, and the blue lines represent the liquid-gas spinodal.}
		\label{fig:Three_Scenarios}
	\end{center}
\end{figure}

{\bf The stability limit conjecture: } The anomalous increase in water's heat capacity and
compressibility with decrease in temperature, with apparent power law divergences at $T_s
= 228 K$ \cite{Speedy_JCP_1976} was explained by Speedy \cite{Speedy_JPC_1982} to be due
to the approach to a spinodal line originating from the liquid-gas critical point. This
spinodal has a positive slope in the (P, T) plane near the critical point, but upon
intersection with the negatively sloped line of density maxima, goes through a zero slope
according to the thermodynamic condition $(dP/dT)_{spinodal} = (\partial P/\partial
T)_{isochore}$ and retraces to higher pressures with a negative slope. Hence in this
scenario the spinodal constitutes both the superheating and supercooling limit of the liquid.
Debenedetti and D'Antonio \cite{DAntonio_JCP_1987, debenedetti_AIChE_1988} further proposed
that thermodynamic consistency also requires that a density maxima locus must necessarily
have an end point. The density maxima locus should either intersect a density minima locus
and hence the liquid shows a normal behaviour in its density or terminate by intersecting at a
spinodal curve (FIG. \ref{fig:Three_Scenarios} (a)). Although some theoretical works have
shown that \cite{Sastry_JCP_1993, Sasai_BCSJ_1993, Borick_JPC_1995} a re-entrant spinodal
is present in models with water like properties, no compelling experimental experimental
verification exists of this scenario \cite{Debenedetti_JPCM_2003}.


{\bf The liquid-liquid critical point scenario: } Poole {\it et al.}
\cite{poole_nature_1992} investigated the retracing spinodal scenario using molecular
dynamics simulations of the ST2 model of water. In their simulation study, the spinodal
was found to be a monotonic function of T. The locus of density maxima or the $TMD$
line, although having a negative slope at high pressures, changes to positive slope at low
pressures (FIG. \ref{fig:Three_Scenarios} (b)). Hence no
intersection between the spinodal and the TMD line occurs. Instead the authors found
evidence for a second critical point, between two forms of the liquid. Considerable
simulation and theoretical investigations since the original work of Poole {\it et al.}
support the possibility of a second critical point. Indeed earlier theoretical analyses
using a two state description \cite{Rapoport_1967_JCP, aptekar_sovphydok_1979} also
generically lead to this possibility \cite{Cuthbertson_PRL_2011}. There has been a
substantial amount of experimental work to verify the possibility of a liquid-liquid
transition in water that has lead to much evidence in support of this possibility,
including recent work on confined water as a way of circumventing crystallisation in bulk
water experiments. Such evidence has been critically reviewed in
\cite{Debenedetti_JPCM_2003, Findenegg_CPC_2008, Holten_JCP_2012}.


{\bf The singularity free hypothesis: } Sastry {\it et al.} \cite{sastry_pre_1996}
proposed that a minimal scenario that was consistent with the salient anomalies did not
require recourse to any thermodynamic singularities, such as a critical point or a
retracing spinodal.  They analysed the interrelationship between the locus of density and
compressibility extrema, and showed that the change of slope of the locus of density
maxima (TMD) was associated with an intersection with the locus of compressibility extrema
($TEC$) (FIG. \ref{fig:Three_Scenarios} (c)). The relation between the temperature
dependence of isothermal compressibility at the TMD and the slope of the TMD is given by

\begin{equation}
	\left ( \frac{\partial K_T}{\partial T} \right )_{P, TMD} = \frac{1}{v}
	\frac{\partial^2 v/\partial T^2}{(\partial P/\partial T)_{TMD}}
\end{equation}

where $K_T$ is the isothermal compressibility. The subscript $P$ and
TMD represents the slope at constant pressure and at the TMD at a given pressure. Since
$\partial^2 v/\partial T^2 > 0$ at the TMD, the above relation shows that for an anomalous
liquid exhibiting a negatively sloped TMD, the isothermal compressibility at constant
pressure increases upon decreasing temperature and hence such increases in compressibility
are not \textit{a priori} an indication of singular behaviour. Calculations with a
lattice model displaying singularity free scenario \cite{sastry_pre_1996, Rebelo_JCP_1998,
LaNave_PRE_1999} reveal a line of compressibility maxima at low temperatures.  The metastable
critical point scenario may be considered to be a special case where the compressibility along the
line of compressibility maxima diverges (at the critical point). Alternatively, it has been argued
that the singularity free scenario is a limiting case where the critical point moves to zero T
\cite{Franzese_PRE_2003}.


{\bf Critical point free scenario: } Recently Angell \cite{angell_Science_2008} has
discussed a possibility, related to some of the early observation of Speedy and Angell
\cite{Speedy_JCP_1976}, in which the high temperature liquid encounters a spinodal at
positive pressure, but this is a spinodal associated with a first order transition between
two liquid states. Such a first order transition however does not terminate in a critical
point, but may terminate at the liquid-gas spinodal. A weaker version of this picture is
that no critical point may exist at positive pressures. Analysis of a model calculation by
Stokely {\it et al.} \cite{Stokely_PNAS_2010} indicates that such a scenario may indeed
arise in the limit of extreme cooperativity of hydrogen bond formation.


\section{Recent Studies of Metastable Silicon}

As described earlier, the early studies of metastable silicon \cite{bagley_AIP_1979,
spaepen_AIP_1979, donovan_APL_1983, Thompson_PRL_1984, Broughton_PRB_1987,
Luedtke_PRB_1989, angell_JNCP_1996} probed the possibility of a liquid-amorphous
transition. More recent work has attempted to find evidence that the transition is one
between two liquid phases.  In this section we present a brief discussion of such recent
work.

\subsection{\bf{Experimental studies}}

Experimental studies of supercooled silicon are very challenging because of high
crystallisation rates. To explore supercooled states by cooling from the high temperature
liquid, one would need to quench the liquid at rates exceeding $10^9Ks^{-1}$
\cite{hedler_nature_2004} to avoid crystallisation and hence using simple quenching
techniques exploring deeply undercooled metastable liquid is not possible. Alternate
methods like chemical vapour deposition and pressure induced techniques
\cite{hedler_nature_2004} have been employed to study the amorphous phase. Other studies
have been performed using methods like aerodynamic levitation \cite{Ansell_JPCM_1998,
Jakse_APL_2003, krishnan_JNCS_2007} or electromagnetic levitation \cite{Higuchi_MST_2005,
Kimura_APL_2001, Watanabe_FarDis_2007, Egry_JNCS_1999, Inatomi_IJT_2007, kim_PRL_2005} to
avoid crystallisation induced by the containers during the experiments.


In silicon, the phase change from a low density liquid ($LDL$) to a high density liquid ($HDL$) involves a change
in electrical conductivity (from a low temperature semiconducting to a high temperature
metallic state), which in turn presents a number of measurable properties that can be used
to detect the phase transition. Optical micrograph methods have been used to measure the
change in optical reflectivity upon a change in phase \cite{McMillan_nmat_2005}, and the
luminescence of the material is also used to detect the phase transition
\cite{Deb_nat_2001}. X-ray diffraction spectra and Raman spectra have also been used to
observe the phase transition \cite{Deb_nat_2001, McMillan_nmat_2005}. Experimental
measurement of densities is quite difficult but {\it in-situ} measurement of structural
quantities and electronic properties have been reported by various groups
\cite{Langen_JCG_1998, Egry_JNCS_1999, rhim_JCG_2000, Sato_IJT_2000, Higuchi_MST_2005,
Higuchi_JNCS_2007, Inatomi_IJT_2007, Watanabe_FarDis_2007, Beye_PNAS_2010}.

\begin{figure}[b]
	\begin{center}
		\includegraphics[scale=0.25]{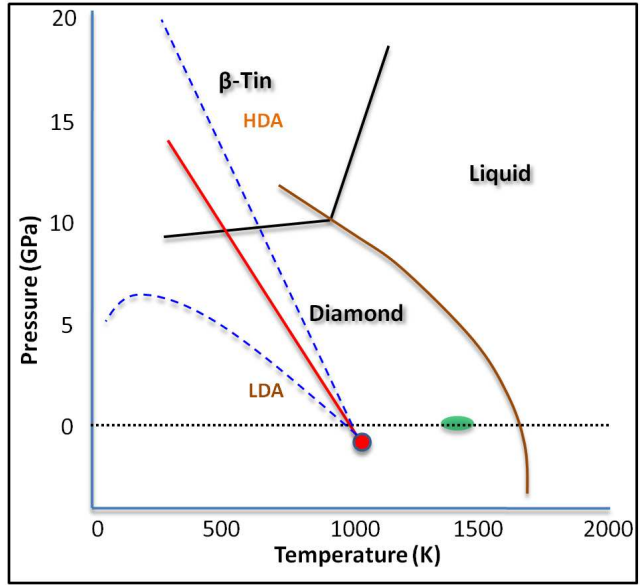}
		\caption{Schematic phase diagram of metastable silicon in the pressure-temperature
		(P, T) plane discussed in \cite{Deb_nat_2001, Daisenberger_PRB_2007}. The brown line
		represents the liquid-crystal (cubic diamond) transition line, extended into the
		$\beta$-Tin phase. The black lines represent the Liquid-$\beta$-Tin and the Cubic
		diamond-$\beta$-Tin transition lines. The red line is the liquid-liquid phase
		transition line ending at a critical point represented by a red circle. The blue
		dotted lines represent spinodals associated with the liquid-liquid transition. The
		green oval represents the amorphous-liquid transition as predicted by some of the
		earlier experiments. [with permission from McMillan\cite{Deb_nat_2001,
		Daisenberger_PRB_2007}.]}
		\label{fig:PD_Deb}
	\end{center}
\end{figure}

Evidence for a pressure induced amorphous-amorphous transition in silicon was first shown
by Deb {\it et al.} \cite{Deb_nat_2001}. These authors studied porous silicon - $\pi$-Si
(silicon having nano-porous holes in its microstructure and a large surface to volume
ratio) because of its luminescence property. At ambient pressure $\pi$-Si exhibits red
luminescence upon irradiating with an Argon laser. With the application of pressure (using
a Diamond anvil cell) the luminescence shifted to longer wavelengths and became opaque at
around $P=10GPa$. X-ray diffraction measurements showed that at around $P=7$ to
$8GPa$ the crystal diffraction pattern disappears and a broad diffraction pattern, a
characteristic of an amorphous material, is observed. At around $P=10$ to $12GPa$, the
crystalline peak disappears entirely. The authors performed Raman scattering
measurements both during compression and decompression and found that upon compression
to $P=13GPa$, the sharp crystalline feature at around wavenumber $520cm^{-1}$
disappears, and a broad peak appears between $200$ to $400cm^{-1}$, distinct from the Raman
signature of tetrahedral low density amorphous ($LDA$) silicon (a broad peak around $400$
to $500 cm^{-1}$). This feature is interpreted to be due to a high density amorphous ($HDA$) phase.
Upon decompression, this feature disappears giving way to a broad peak around
$400$ to $500 cm^{-1}$, corresponding to LDA at low pressure. These observation led to the
conclusion that $\pi$-Si undergoes a pressure induced amorphous-amorphous phase
transition. In turn, this amorphous-amorphous transition was suggested to be related to a
liquid-liquid transition, employing a theoretical model \cite{Moynihan_JNCS_2000}. The
schematic phase diagram of metastable silicon (extracted from the work of Deb {\it et al.}
\cite{Deb_nat_2001}) is shown in FIG. \ref{fig:PD_Deb}.

\begin{figure}[t]
	\begin{tabular}{cc}
		\subfloat[]{\includegraphics[width=0.38\textwidth]{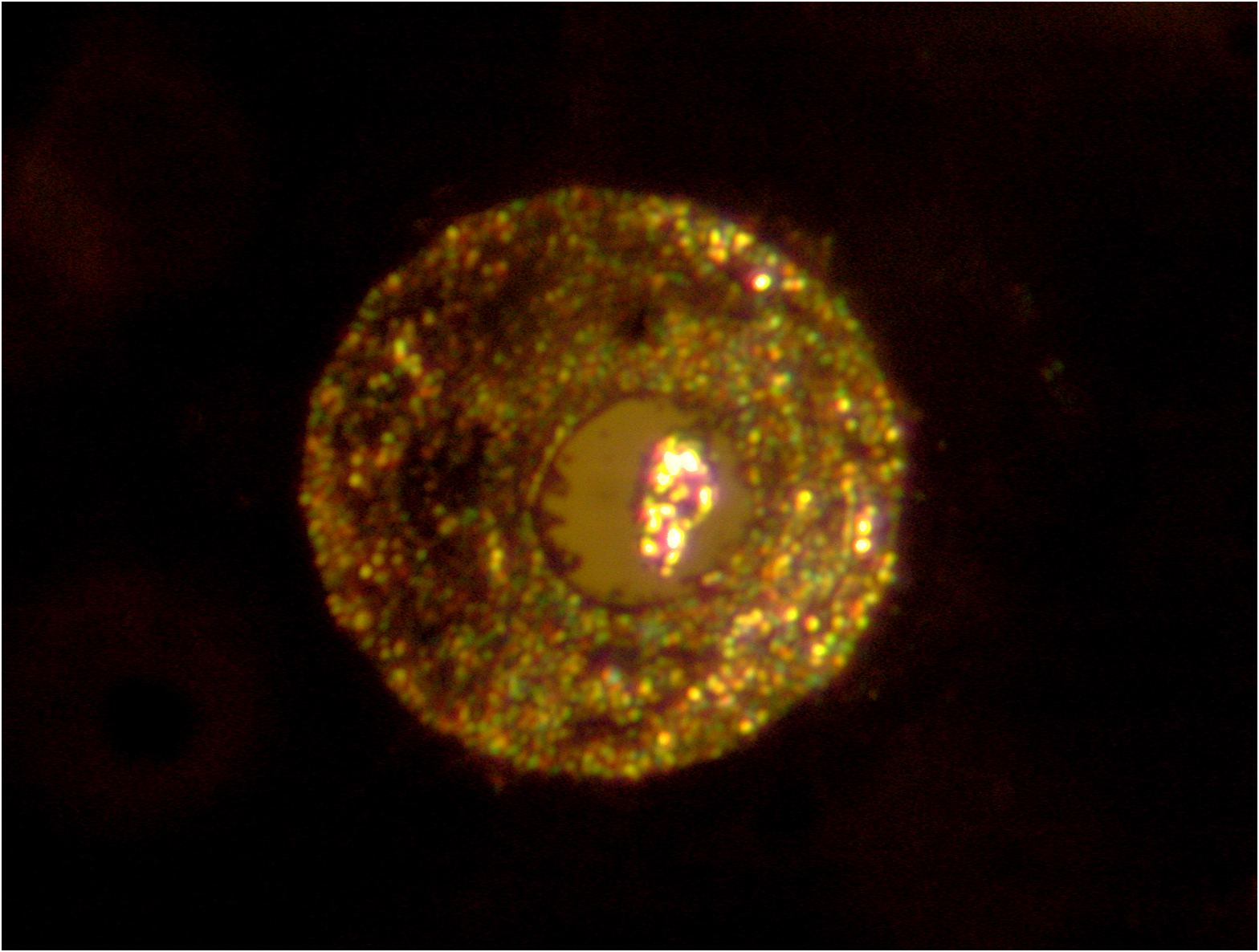}}\hspace{5 mm}
		&\subfloat[]{\includegraphics[width=0.373\textwidth]{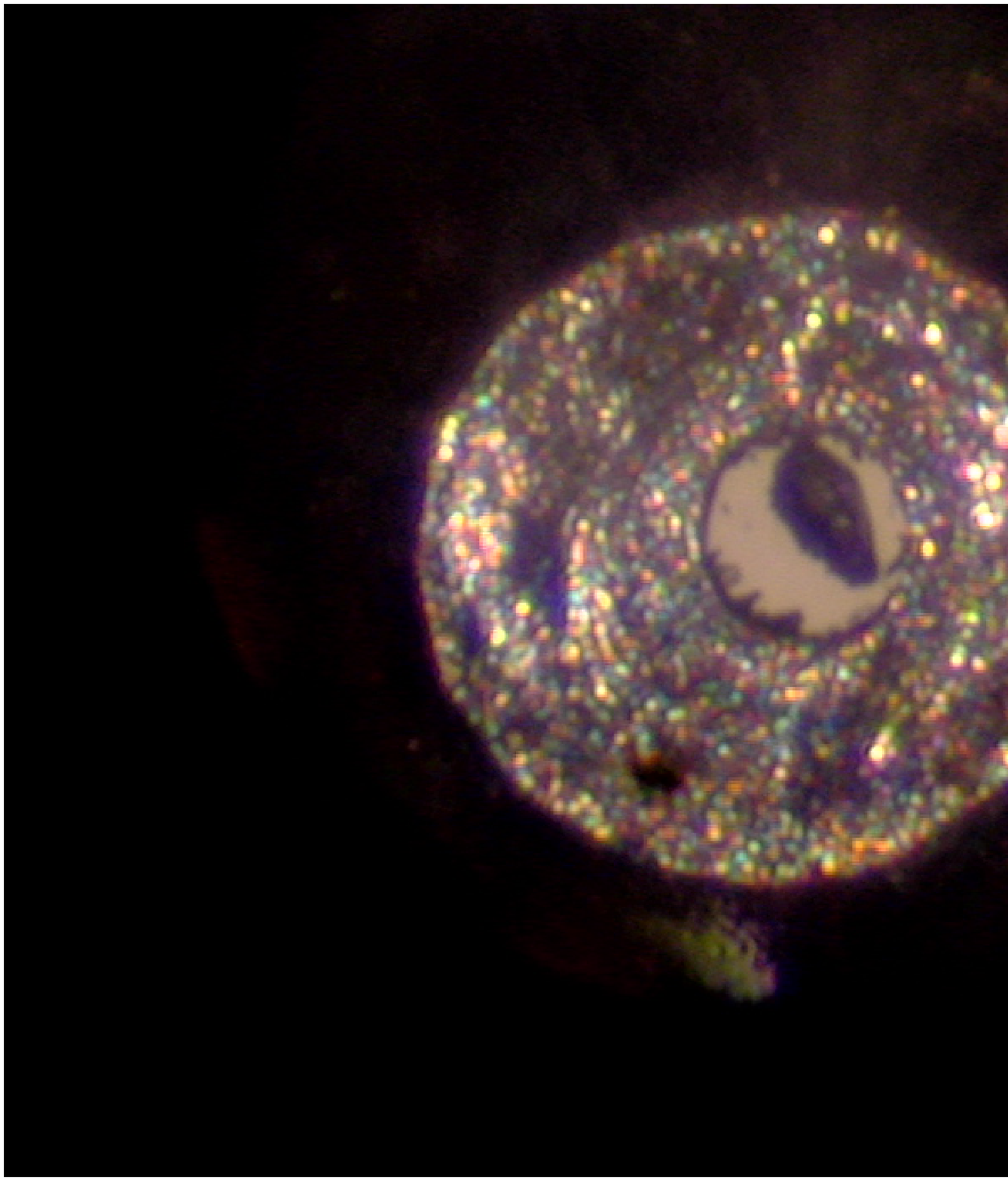}}\\
	\end{tabular}
	\caption{Optical micrographs of an amorphous silicon sample show that HDA at $P = 16.6 GPa$
	(left) is highly reflective and LDA at $P = 13.5 GPa$ (right) is non-reflective (compared to
	the surrounding metal gasket). [with permission from McMillan and Daisenberger
	\cite{McMillan_nmat_2005, Daisenberger_JPCB_2011}.]}
	\label{fig:aSi_OpObs}
\end{figure}

Direct optical observation and electrical resistance measurements carried out on amorphous silicon by
McMillan and co-workers \cite{McMillan_nmat_2005} showed that the HDA is highly reflective
and LDA is non-reflective ({\it see} FIG. \ref{fig:aSi_OpObs}). From the electrical resistance
measurements the authors found that there is an abrupt decrease in resistivity across the
LDA-HDA transition around $P = 11 GPa$, indicating transformation to metallic
HDA. The sample was verified to be in its amorphous state (using Raman spectroscopy),
since pressure induced crystallisation to $\beta$-Sn phase could also lead to a drop in
resistivity.


The above experiments suggest the possibility of a liquid-liquid transition but are performed under
conditions at which the amorphous forms of silicon are solid. In an attempt obtain a more direct
evidence that the transition is between two liquids, Hedler {\it et al.}
\cite{hedler_nature_2004} performed ion bombardment experiments on amorphous silicon. The plastic
deformations they observe of the samples are similar to the deformation seen in
conventional glasses undergoing the glass transition, and the authors deduce a glass
transition of around $1000 K$.


As described below, in both classical and {\it ab initio} simulations a clear evidence of
liquid-liquid phase transition has been found. Computer simulations
\cite{sastry_nmat_2003, Miranda_JCP_2004, ashwin_prl_2004, Jakse_APL_2003,
Ganesh_PRL_2009, Vasisht_nphys_2011, Garcez_JCP_2011} predict that the first order phase
transition is characterised by a change in coordination number from $4$, in LDL to greater
than $5$, in HDL. It has been also observed that the tetrahedrally coordinated LDL is less
diffusive compared to HDL. The electronic structure calculations \cite{ashwin_prl_2004,
Jakse_APL_2003, Ganesh_PRL_2009} in these simulations have shown that the LDL is less
metallic than HDL.

\begin{figure}[b]
	\begin{center}
		\includegraphics[scale=0.5]{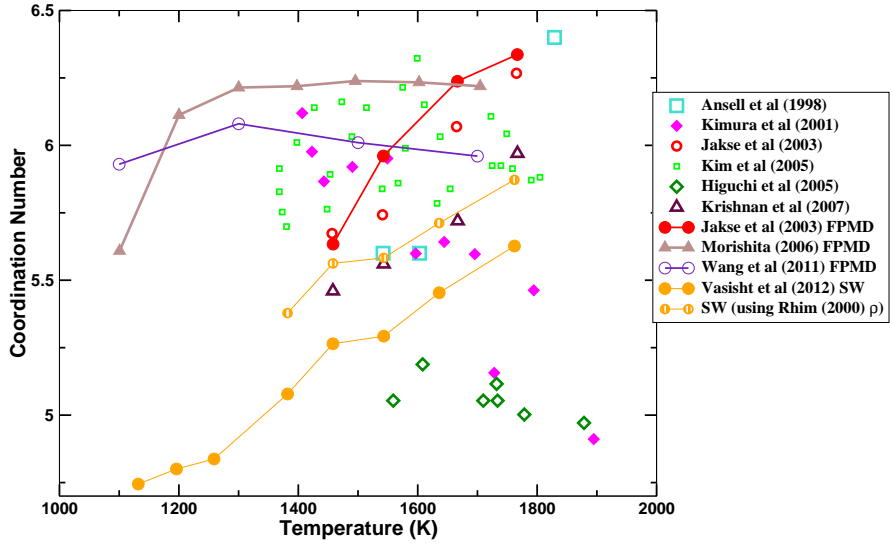}
		\caption{Compilation of coordination number measurements plotted against temperature
		(at $P=0GPa$) as reported by different experimental reports, first principle MD
		(FPMD) simulations as well as classical simulations results. [From Ansell {\it et
		al.} \cite{Ansell_JPCM_1998}, Kimura {\it et al.} \cite{Kimura_APL_2001}, Jakse {\it
		et al.} \cite{Jakse_APL_2003}, Kim {\it et al.} \cite{kim_PRL_2005}, Higuchi {\it et
		al.} \cite{Higuchi_MST_2005}, Krishnan {\it et al.} \cite{krishnan_JNCS_2007},
		Morishita \cite{Morishita_PRL_2006}, Wang {\it et al.} \cite{Wang_PBCM_2011} with
		permission.]}
		\label{fig:Cnn}
	\end{center}
\end{figure}

Different Experimental groups have tried to measure the coordination number seeking
evidence for liquid-liquid phase transition. To circumvent the container induced
crystallisation these experiments are carried out by levitating the sample. The
coordination number obtained from different experimental reports are put together in FIG.
\ref{fig:Cnn}. With the state of the art in experimental techniques, the lowest
temperature achievable, keeping the sample in liquid state, is around
$T=1380K$ \cite{kim_PRL_2005}. As it can be seen from FIG. \ref{fig:Cnn}, there is quite a
large spread in the coordination numbers as calculated from different experimental
groups. Secondly in the measured range of temperatures the coordination number remains
greater than $5$. These results do not agree with the predicted liquid-liquid phase
transition temperature at ambient pressure from earlier experimental works, and this issue
remains one that needs to be properly understood. A possible explanation is that the
coordination number is quite sensitive to the density of the liquid (a quantity which is
difficult to measure in experiments). Another possibility is that indeed the previous
estimates of the transition temperature are high.

Recently, Beye {\it et al.} \cite{Beye_PNAS_2010} used femto-second pump-probe
spectroscopy, and the expected changes in the electronic structure of silicon to attempt a
direct verification of the liquid-liquid transition, by monitoring the evolution of
electronic density of state ($DOS$). After exciting a sample of the crystal with a pump
X-ray pulse, they monitored the evolution of the electronic DOS, and found it to evolve in
a two step process, with the intermediate step showing clear resemblance to the DOS of
LDL, and the DOS at later times resembling that of HDL. Although the process during which
these measurements are made are highly non-equilibrium in nature and there are gaps in our
understanding, these results point the way to how direct evidence for the liquid-liquid
transition under extreme metastable conditions may be obtained experimentally.


\subsection{\bf{Simulation studies: Phase behaviour, structure and dynamics}}

\subsubsection{\bf{Liquid-liquid transition at zero pressure}}

\begin{figure}[t]
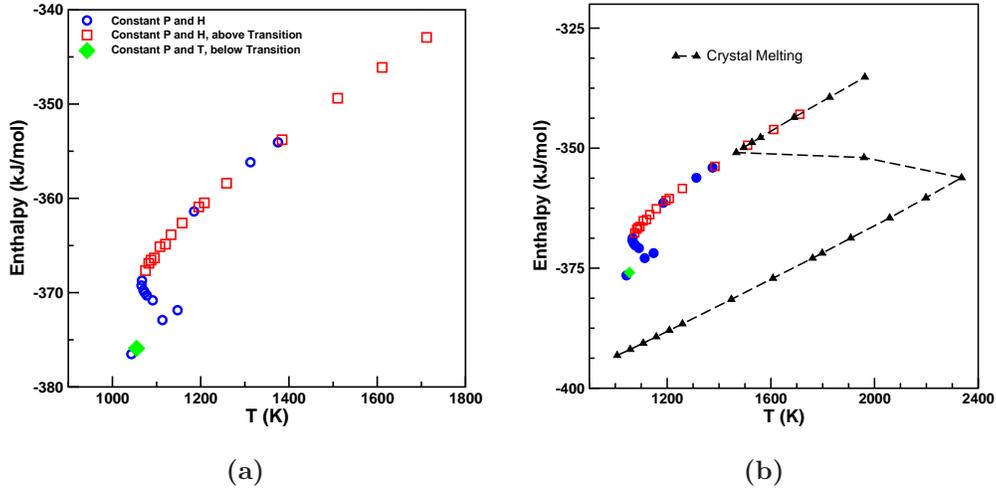

	\begin{tabular}{cc}
		\subfloat[]{\includegraphics[width=0.385\textwidth]{images/Recent_Work/Simulation/SA_Latentheat.eps}}\hspace{5 mm}
		&\subfloat[]{\includegraphics[width=0.375\textwidth]{images/Recent_Work/Simulation/SA_Latentheat_Cry.eps}}\\
	\end{tabular}
	\caption{(a) The enthalpy against temperature from NPH MD simulations and NPT MD simulations
	using the SW potential for the supercooled liquid above and below the liquid-liquid
	transition. (b) The crystal-liquid transition is shown for comparison with the liquid-liquid
	transition data. [From Sastry {\it et al.}\cite{sastry_nmat_2003} with permission.]}
	\label{fig:SA_Latentheat}
\end{figure}

Simulation work described up to now, although supporting a first order transition to a low
density liquid upon cooling below $T=1060 K$, are subject to uncertainties of
interpretation owing to the low mobility of the low temperature states which did not
permit an unambiguous demonstration of a first order {\it liquid-liquid} transition. The
simulation study of Sastry and Angell \cite{sastry_nmat_2003} addressed these
uncertainties, by seeking evidence of (a) phase co-existence, and (b) finite mobility in
the low temperature phase.  To probe whether a first-order transition exists, the authors
carried out constant enthalpy (NPH) simulations. A non-monotonic dependence of the
enthalpy on temperature was found (FIG. \ref{fig:SA_Latentheat}), which is an indication
of a first order phase transition. The transition temperature was found to be around
$T=1060 K$ at zero pressure. Similar behaviour is also observed in first principles
simulations by Jakse and Pasturel \cite{Jakse_PRL_2007}.

Further, the authors studied the nature of two phases, by looking at their structural and
dynamic properties. The mean square displacement ($MSD$) obtained from constant pressure
simulations on either side of the phase transition showed a linear behaviour with time,
indicating that the phases are in the liquid state with finite diffusivity. The
diffusivity values calculated from MSD at various temperatures showed roughly a two orders
of magnitude drop as the high temperature liquid transforms into the low temperature liquid (FIG.
\ref{fig:SA_Diffusive}). The equilibration times in the low temperature liquid phase (see
below) range from tens to hundreds of nanoseconds. The T-dependence of the diffusivity in
the high temperature liquid phase (till $T=1070K$) was found to be highly non-Arrhenius,
characteristic of a {\it fragile liquid} \cite{Sastry_PhyA_2002}.

\begin{figure}[t]
	\begin{center}
		\includegraphics[scale=0.35]{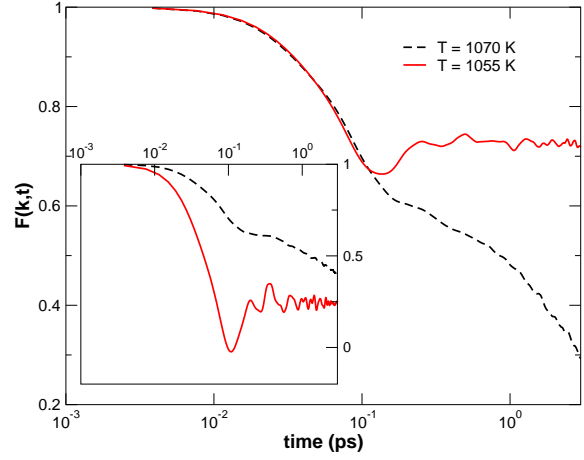}
		\caption{ {\it Main panel}: The intermediate scattering function F(k,t) from MD
		simulations using the SW potential of 512 particles, above and below the transition.
		The low temperature liquid displays damped oscillatory behaviour, characteristic of
		strong liquids. The high temperature liquid shows a monotonic decrease,
		characteristic of fragile liquids. {\it Inset}: The intermediate scattering function
		for smaller system size (108 particles). [From Sastry {\it et al.}
		\cite{sastry_nmat_2003} with permission.]}
		\label{fig:SA_InterScatt}
	\end{center}
\end{figure}

It has been argued \cite{Angell_PCCP_2000} that silicon, along with other tetrahedral
liquids such as water and silica, should undergo a transition from fragile liquid behaviour
(non-Arrhenius temperature dependence of viscosity and other transport coefficients) to
strong liquid behaviour (Arrhenius temperature dependence) as the liquid makes a transition
from the HDL to the LDL (either discontinuously or
continuously at pressures below the critical pressure). Since data over a sufficient range
of temperatures in the low temperature phase was not available to judge this matter
directly for silicon, Sastry and Angell \cite{sastry_nmat_2003} took recourse to an
empirical observation that the intermediate scattering function (F(k,t)) shows oscillatory
behaviour in strong liquids that becomes more pronounced in small systems. In HDL phase
($T=1055 K$) no oscillations were observed in the F(k,t) ({\it see} FIG.
\ref{fig:SA_InterScatt}) and upon transition to LDL phase ($T=1070K$) oscillatory
behaviour appears and it becomes more significant at lower system sizes ({\it inset} of FIG.
\ref{fig:SA_InterScatt}). Independent evidence for such a transition also is obtained by
the fact that the heat capacity $C_P$ drops to a value of $3.6 Nk_B$ in the low
temperature phase \cite{Jakse_JCP_2009}.

\begin{figure}[b]
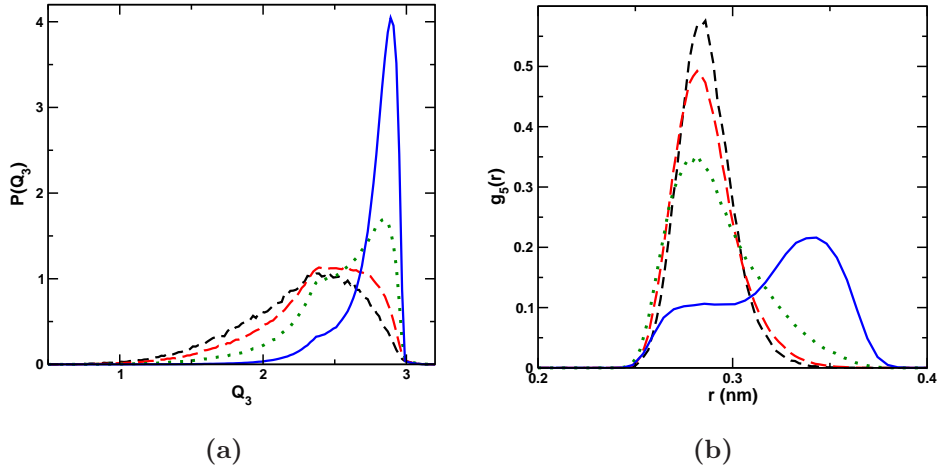

	\begin{tabular}{cc}
		\subfloat[]{\includegraphics[width=0.348\textwidth]{images/Recent_Work/Simulation/Q3.eps}}\hspace{5 mm}
		&\subfloat[]{\includegraphics[width=0.365\textwidth]{images/Recent_Work/Simulation/g5ofr.eps}}\\
	\end{tabular}
	\caption{(a) The distribution of local bond orientation order parameter ($Q_3$) from MD
	simulations using the SW potential. The continuous blue line is for the low temperature
	liquid, which indicates local tetrahedral ordering. (b) The fifth neighbour distance
	distribution $g_5(r)$. For the high temperature liquid (dotted lines), $g_5(r)$ show a
	uni-modal peak indicating that the fifth neighbour is within the first coordination shell.
	For the low temperature liquid (continuous blue line), a bimodal distribution emerges
	indicating the expulsion of of the fifth neighbour in a majority of cases to distances
	outside the first coordination shell. [From Sastry {\it et al.} \cite{sastry_nmat_2003} with
	permission.]}
	\label{fig:SA_Q3_g5ofr_Dist}
\end{figure}

The pair correlation function $g(r)$, fifth neighbour distribution $g{_5}(r)$, bond angle
correlation function $G_3$ and local bond orientation order parameter $Q_3$ (see $IV B (4)$ for
definitions were calculated to study the structural properties of the two liquids. Coordination
numbers calculated by integrating the $g(r)$ till its first minimum. The average coordination
number was found to change from $5.12$ to $4.61$ in the high-temperature liquid and around
$4.2$ in the low-temperature liquid. The fraction of four coordinated atoms increased from
about $50\%$ (at high T) to $80\%$ in the low T phase, indicating a greater degree of
local tetrahedral order. This change is  also reflected in the local bond orientation order
$Q_3$({\it see} FIG. \ref{fig:SA_Q3_g5ofr_Dist} (a)). The $Q_3$ values for the low temperature
liquid peaked at the crystal's $Q_3$ value, suggesting a tetrahedral local ordering whereas the high
temperature liquid showed a broader peak in $Q_3$. The fifth neighbour distribution (which is
distribution of distances of the $5^{th}$ nearest neighbour to a given atom) for the high temperature
liquid was found to be unimodal indicating that the fifth neighbour resides inside the first
coordination shell, whereas the low temperature liquid showed a bimodal $g_5(r)$
distribution ({\it see} FIG. \ref{fig:SA_Q3_g5ofr_Dist} (b)), with the larger peak shifting to the second
neighbour shell in the low temperature phase.



\subsubsection{\bf{Liquid-liquid critical point}} 

Recently, extending the work of Sastry and Angell \cite{sastry_nmat_2003}, which provided
evidence of a first order liquid-liquid phase transition in silicon at zero pressure,
Vasisht {\it et al.} \cite{Vasisht_nphys_2011, Vasisht_nphys_sup_2011} reported evidence
of the existence of a negative pressure liquid-liquid critical point in SW silicon, based
on extensive simulations of the SW model of silicon. These authors also constructed the
complete phase diagram of supercooled silicon, which clearly demonstrates the
interconnection between various thermodynamic anomalies and the phase behaviour of the
liquid as analysed in previous works \cite{Speedy_JCP_1976, Speedy_JPC_1982,
debenedetti_AIChE_1988, poole_nature_1992, sastry_pre_1996, Rebelo_JCP_1998, poole_JPCM_2005}.

\begin{figure}[t]
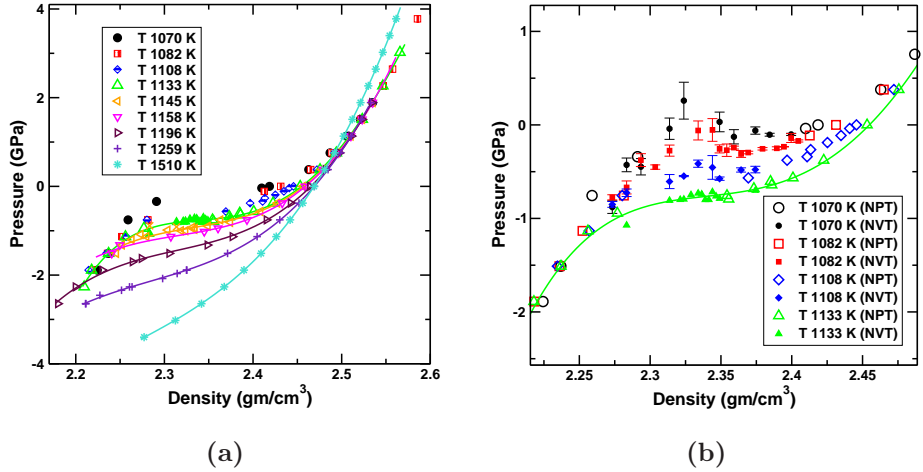

	\begin{tabular}{cc}
		\subfloat[]{\includegraphics[width=0.35\textwidth]{images/Si_LLPT/EOS_Above_Tc_Real.eps}}\hspace{5 mm}
		&\subfloat[]{\includegraphics[width=0.35\textwidth]{images/Si_LLPT/EOS_Below_Tc_Real_New.eps}}\\
	\end{tabular}
	\caption{Equation of state from NPT MD and NVT MD simulations using the SW potential. Nine isotherms at
	temperatures above and below the critical temperature of the liquid-liquid transition are
	shown. The open symbols represent data from NPT MD simulations and the opaque symbols
	represent data from NVT MD simulation. The solid lines are polynomial fits to the data
	points. (a) The isotherms above $T = 1133K$ are monotonic and continuous and below $T =
	1133K$ show a jump in density for small change in pressure in constant pressure simulations.
	(b) Constant volume (NVT) MD simulation data for $T < 1133 K$ show non-monotonic behaviour
	indicating a first order phase transition.}
	\label{fig:EOS_Above_Below}
	\vspace{5mm}
\end{figure}

It is well known that the liquid-gas coexistence line does not extend
to arbitrarily large T, but terminates at the {\it critical point},
where second derivatives of the free energy are singular ({\it e. g.},
the heat capacity and the isothermal compressibility). Density along
isotherms {\it vs.} pressure are continuous at temperatures above the
critical temperature, whereas they display a density discontinuity at
lower temperatures, at the coexistence pressure. The same kind of
behaviour is expected for the liquid-liquid transition, and hence,
Vasisht {\it et al.} computed the equation of state (pressure {\it
  vs.} density for varying temperature) in order to study the phase
behaviour \footnote{A summary of the details of the simulations
  employed in \cite{Vasisht_nphys_2011} are as follows: Constant
  pressure (NPT) molecular dynamics (MD) simulations using SW
  potential were performed with a time step of 0.383 fs and a system
  size of 512 particles, employing an efficient algorithm
  \cite{Saw_JCP_2011} for energy and force evaluations. Constant
  volume (NVT) simulations were performed with the same system size
  and time step using the LAAMPS \cite{Plimpton_JCP_1995} parallelised
  MD package. In the HDL phase, a minimum of $3$ to $6$ independent
  samples were simulated for $\sim 100$ relaxation times ($\sim$ $10$
  $ns$). In the LDL phase, since the authors find that the
  crystallisation (monitored by energy jumps, mean square displacement
  (MSD) and pair correlation function) rates are high, around $10$ to
  $50$ initial runs were performed, each of $22$ ns. Noncrystallising
  samples (average of $5$) were run up to $10$ relaxation times when
  possible and in all LDL cases, simulations were carried out for
  times required for the MSD to reach $1 nm^2$ (5 $\sigma^2$, where
  $\sigma$ is the atomic diameter) or for $100 ns$ ($300$ million MD
  steps), whichever was larger. Equilibration of the system was
  monitored by the MSD and from the relaxation of self overlap
  function Q(t) defined as $Q(t) = {1\over N}\sum_{i=1}^{N} w \left
  |\vec r_i(t_0) - \vec r_i(t + t_0) \right |$, where $w(r)$ = $1$, if
  $r \leq 0.3\sigma$, zero otherwise. Parallel tempering Monte Carlo
  (MC) simulations \cite{Sever_JPCA_2001} are employed to equilibrate
  the system at very low temperature and high negative pressure (while
  deducing the temperature of minimum density) and restricted ensemble
  MC simulations for locating the spinodal at low
  temperatures.}. Vasisht {\it et al.}\cite{Vasisht_nphys_2011}
constructed the equation of state (EOS) of supercooled silicon for
temperatures ranging from $T=1070 K$ to $1500 K$. From the EOS the
upper and lower bound in temperature for the critical point was
found. Above the critical temperature the EOS is a continuous and
monotonic curve as shown in the FIG. \ref{fig:EOS_Above_Below} (a). Below the critical temperature, the system phase separates and
hence these isotherms (below $T = 1133 K$), in the isothermal-isobaric
$(NPT)$ MD simulations, showed jumps in densities for a small change
in pressure (FIG. \ref{fig:EOS_Above_Below} (b)) The
isothermal-isochoric $(NVT)$ MD simulations performed at these
temperatures ($T = 1133 K$ to $1070 K$) (and at densities spanning the
range of the density jumps in $(NPT)$ MD simulations) showed
non-monotonic isotherms (FIG. \ref{fig:EOS_Above_Below} (b)). Such non-monotonicity in simulations arises from metastability
on the one hand, and on the other hand, incomplete phase separation
owing to finite sample sizes in the unstable region, and constitutes a
clear indication of a first order transition. The highest
phase-coexisting temperature and lowest continuous isotherm bounds the
critical temperature. Similarly, the pressure at which the continuous
isotherms are flattest (above the critical temperature) and the
pressure at which a density jump is seen (below the critical
temperature) determine the bounds for the critical pressure. The
estimated critical temperature and pressure are $T_c \sim 1120 \pm 12
K$, and $P_c \sim -0.60 \pm 0.15 GPa$.

\begin{figure}[b]
	\begin{center}
		\includegraphics[scale=0.45]{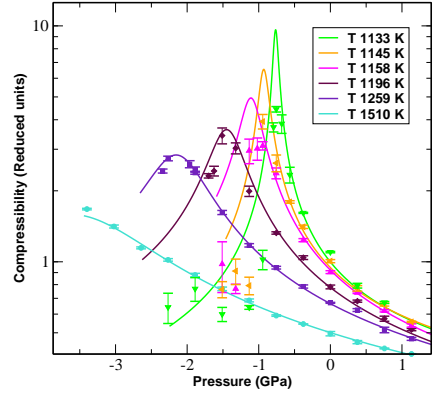}
		\caption{Isothermal compressibility against pressure for different temperatures from
		NPT MD simulations using the SW potential. All the isotherms shown in the  figure are
		for temperatures above the liquid-liquid critical temperature. With the decrease in
		temperature the maximum value of the compressibility along an isotherm increases,
		suggesting an approach to the critical point. The lines represent the
		compressibility values calculated from the equation of state by numerical
		differentiation. The symbols represent the compressibility calculated from volume
		fluctuations.}
		\label{fig:Compressibility}
	\end{center}
\end{figure}

Approaching the critical point from above leads to increased density fluctuations.
Compressibility value were calculated from density fluctuations (in NPT MD
simulation), in addition to evaluating the $K_T$ from the equation of state (EOS). $K_T$
from the equation of state was calculated by taking the derivative of the equation of
state (after doing a polynomial fit to the data points obtained from the NPT simulation)
and from volume fluctuations using the relation $K_T = \frac{1}{k_{B} T}\frac{<V^{2}> - <V>^{2}}{<V>
}$.  In FIG. \ref{fig:Compressibility}, the compressibility values calculated from both
methods for temperatures above $T = 1133 K$ are shown. As seen in FIG.
\ref{fig:Compressibility} the EOS estimates for agree well with those from density
fluctuations in the HDL, but poorer agreement is obtained for the LDL, because of poorer sampling.

\begin{figure}[t]
	\begin{center}
		\includegraphics[scale=0.25]{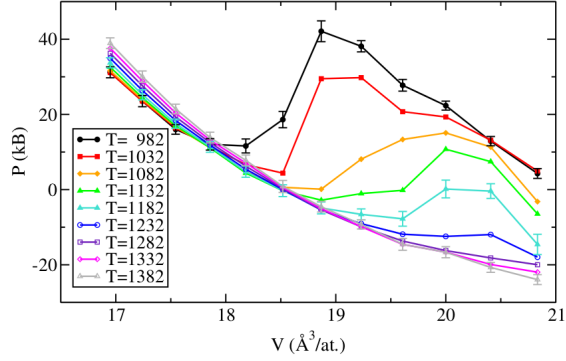}
		\caption{Equation of state of supercooled silicon obtained from first principles MD
		(FPMD) simulations displaying a van der Waals-like loop for $T < 1232K$. [From
		Ganesh {\it et al.} \cite{Ganesh_PRL_2009} with permission.]}
		\label{Ganesh_EOS}
	\end{center}
\end{figure}

\textbf{First principles simulations: } In the cases of carbon and silica, computer
simulations using classical empirical potentials have shown a liquid-liquid transition
\cite{Glosli_PRL_1999, Ivan_PRE_2000}, but first principle MD ($FPMD$) simulations
\cite{Wu_PRL_2002, karki_PRB_2007} show results that are not consistent with
classical simulations. In silicon, Jakse and Pasturel \cite{Jakse_JCP_2008} and
independently Ganesh and Widom \cite{Ganesh_PRL_2009} have reported first principle
simulation results, both of which support the proposed liquid-liquid transition in
silicon. In the work of Ganesh and Widom, the authors report the emergence of a van der
Waals-like loop (shown in FIG. \ref{Ganesh_EOS}), as signature of a first order phase
transition at temperatures below $1182 K$. The maximum time span of these
simulations is around $40ps$ \cite{Jakse_JCP_2008}, which seems to be very small
compared to the relaxation times of LDL (tens to hundreds of nanoseconds; see below)
obtained from simulations of SW silicon \cite{sastry_nmat_2003}. But the FPMD calculations
are computationally very expensive compared to classical MD simulations. Hence it would be
of interest to compare the equilibration times of the system simulated in FPMD and
classical MD and also do a systematic study of relaxation processes in these two different
methods of simulation. A comparison of properties obtained in different simulations are
discussed in a later section.

\subsubsection{\bf{Phase diagram}}

\begin{figure}[b]
	\begin{center}
		\includegraphics[scale=0.5]{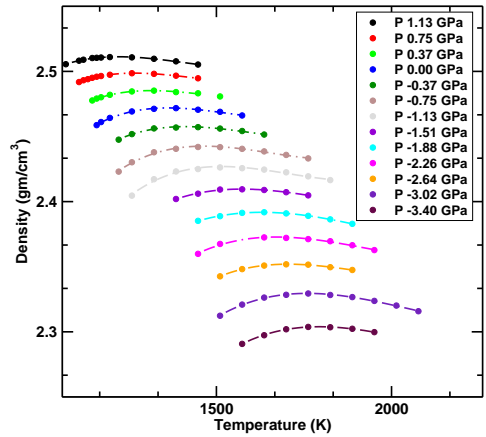}
		\caption{Density against temperature for different isobars from NPT MD simulations
		using the SW potential. The temperature of the maxima along the isobars as a
		function of the pressure defines the TMD line.}
		\label{fig:TMD_Isobar}
	\end{center}
\end{figure}

\begin{figure}[t]
	\begin{center}
		\includegraphics[scale=0.6]{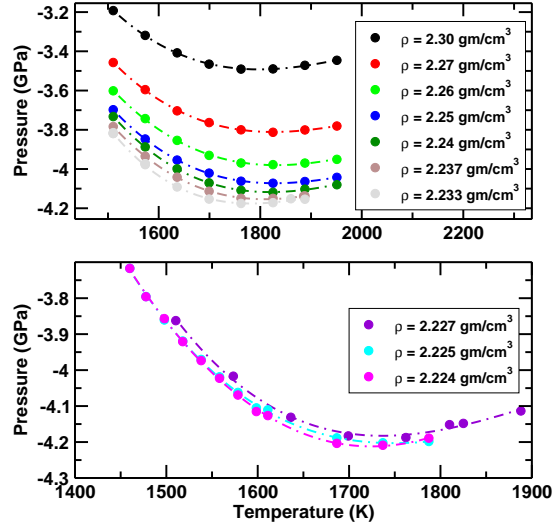}
		\caption{\textit{(top panel)} Pressure against temperature for different
		isochores from NVT MD simulations using the SW potential. The pressure and
		temperature values at the minimum obtained along each isochore for varying density
		define the TMD line in the (P,T) plane. \textit{(bottom panel)} Isochores obtained
		from NVT MD simulations at the lowest three densities. Below these densities, the
		system cavitates before the isochore passes through a minimum.}
		\label{fig:TMD_Isochore}
	\end{center}
\end{figure}

In order to obtain the complete phase behaviour of supercooled silicon Vasisht {\it et
al.} analysed the interplay of various loci of extremal behaviour, namely the spinodal,
temperature of density extrema and temperature of compressibility extrema. Loci of
temperature of maximum and minimum density (TMD and $TMinD$), temperature of maximum and
minimum compressibility ($TMC$ and $TMinC$) and spinodal were evaluated by employing, in
addition to the MD simulations, parallel tempering ($PT$) Monte Carlo simulations (at low
temperature and pressures) and restricted ensemble Monte Carlo ($REMC$) simulations
\cite{corti_CES_1994} (for locating the spinodal at low temperatures). Results concerning
these features of the phase diagram are described below:

{\bf Temperature of Maximum Density: }The TMD line is defined as the locus of isobaric
maxima of density $\rho$ {\it vs.} $T$ ($(\partial \rho/\partial T)_P = 0$) or the locus
of isochoric minima of pressure $P$ {\it vs.} $T$ ($(\partial P/\partial T)_V = 0$). For
pressure values above $P = -3.80 GPa$, the TMD line was obtained from NPT MD
simulations. Below $P = -3.80 GPa$, cavitation in NPT MD simulations was observed and hence
NVT MD simulations were performed to locate isochoric minima of pressure. The TMD obtained
from density maxima along isobars and pressure minima along isochores are shown in FIG.
\ref{fig:TMD_Isobar} and  FIG. \ref{fig:TMD_Isochore} respectively.

\begin{figure}[b]
	\begin{center}
		\includegraphics[scale=0.6]{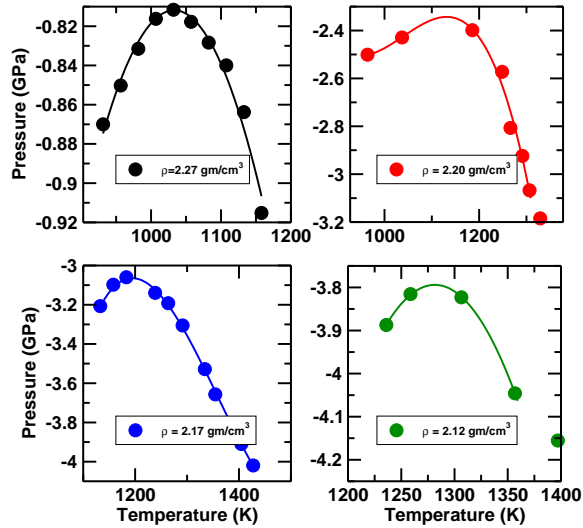}
		\caption{Pressure against temperature for different isochores from parallel
		tempering MC simulations using the SW potential. The location of the maxima along
		the isochores define the TMinD line.}
		\label{fig:TMinD_Isochore}
	\end{center}
\end{figure}

{\bf Temperature of Minimum Density: }The TMinD line is locus of density minima points,
crossing which, the system returns to normal behaviour (density increases with decrease in
temperature). Finding the TMinD line in supercooled silicon is challenging since one must
simulate the system deep inside the supercooled region of the phase diagram (where
crystallisation, slow equilibration and cavitation pose hurdles to obtaining equilibrated
data). In order to obtain equilibrated data NPT parallel tempering MC simulation technique
was employed \cite{Sever_JPCA_2001}, in which copies of the system at different T and P
are swapped periodically according to a Metropolis acceptance criterion, thereby avoiding
the possibility of the system getting stuck in phase space at low temperatures. The TMinD
obtained from maxima along isochores are shown in FIG. \ref{fig:TMinD_Isochore}. There
have been very few reports of density minima for any substance. Experimental and
simulation observation for water were reported only recently\cite{poole_JPCM_2005,
liu_PNAS_2007}.

\begin{figure}[t]
	\begin{center}
		\includegraphics[scale=0.6]{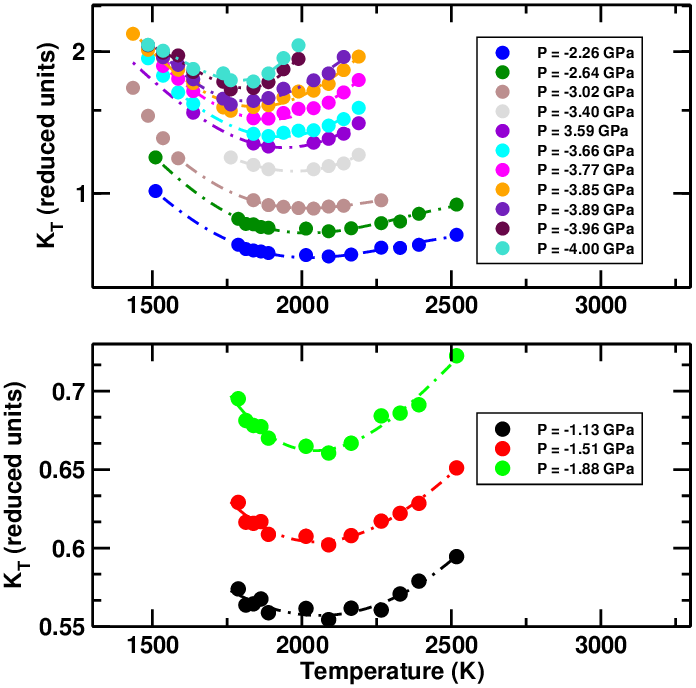}
		\caption{Isothermal compressibility against temperature for different isobars from
		MD simulations using the SW potential. The location of the minima along the isobars
		define the TMinC line.}
		\label{fig:TMinC_Isobar}
	\end{center}
\end{figure}

{\bf Temperature of Minimum Compressibility:} Using NPT MD simulations the line of TMinC
was obtained ({\it see} FIG. \ref{fig:TMinC_Isobar} {\it top panel}). At pressure values below $P =
-3.80 GPa$ system cavitates quite easily and often.  Hence care was taken by performing
simulations for a minimum of 10 independent samples to construct the equation of state
(EOS), from which the compressibility was calculated.
        
\begin{figure}[b]
	\begin{center}
		\includegraphics[scale=0.6]{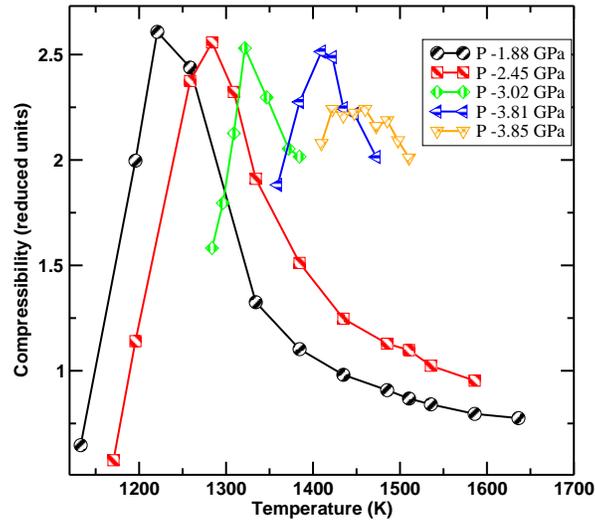}
		\caption{Isothermal compressibility against temperature for different isobars from MD
		simulations using the SW potential. The location of the maxima along the isobars
		define the TMC line.}
		\label{fig:TMaxC_Isobar}
	\end{center}
\end{figure}

{\bf Temperature of Maximum Compressibility:} The value of TMC at the high pressure values
($P > -2 GPa$) is obtained from the FIG. \ref{fig:Compressibility}. Compressibility data
from which $K_T^{max}$ are obtained for $P < -2 GPa$ are shown in the  FIG.
\ref{fig:TMaxC_Isobar}. As the system crosses the $K_T^{max}$ line from high T to low T
(at a chosen pressure value), the relaxation times were found increase from picoseconds to
tens of nanoseconds. Nearing the compressibility maxima,  crystallisation of samples was
also found to be frequent. The $K_T$ values shown in the FIG. \ref{fig:TMaxC_Isobar} are
calculated from both volume fluctuations measured in NPT MD simulations and from
derivatives of pressure from NVT MD simulations. For pressure values below $-3.90 GPa$, the
system cavitates easily and hence the location of $K_T^{max}$ at these state points were
not evaluated.

\begin{figure}[t]
	\begin{center}
		\includegraphics[scale=0.6]{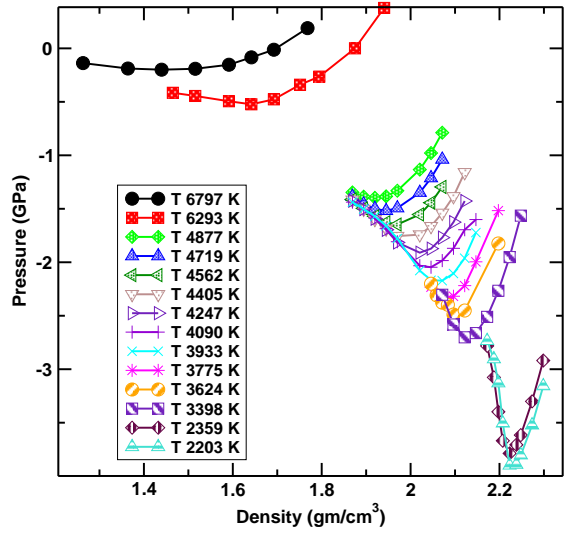}
		\caption{Pressure against density for high temperature isotherms ($T > 2200 K$)
		from NPT MD simulations using the SW potential. The location of the minima along the
		isotherms define the spinodal line.}
		\label{fig:HighT_Spinodal}
	\end{center}
\end{figure}

{\bf Liquid spinodal:} A spinodal point is defined by the condition $(\partial P/\partial
V)_T = 0$. FIG. \ref{fig:HighT_Spinodal} shows high temperature spinodal isotherms ($T >
2200 K$). These isotherms were obtained from constant volume temperature (NVT) MD
simulations.

\begin{figure}[b]
	\begin{center}
		\includegraphics[scale=0.6]{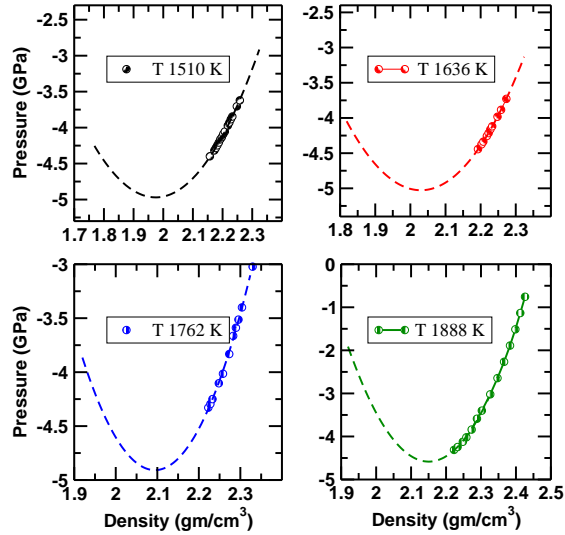}
		\caption{Pressure against density for low temperature isotherms ($T < 2200 K$) from
		MD simulations using the SW potential. The dashed line indicate the quadratic
		extrapolation of the form $p_0 + a1\times(\rho-\rho_0) + a2\times(\rho-\rho_0)^2$ which 
are used to locate the spinodal.}
		\label{fig:LowT_Spinodal}
	\end{center}
\end{figure}

For T $<$ $2200$ K, cavitation was observed in the NVT simulations before the minima along
the isotherms are reached, due to which a drastic increase in the pressure was observed.
In an attempt to circumvent this problem, the authors performed REMC simulations wherein
arbitrary bounds were imposed on the magnitude of the allowed density fluctuations by
dividing the simulation box into a number of equal sub-cells and constraining the number
of particles in each of these sub-cells \cite{corti_CES_1994}. However, even in the REMC
simulations, the system was found to cavitate occasionally, with the formation of voids
across sub-cell boundaries (with each sub-cell satisfying the applied constraint on number
of particles). Hence the estimation of the spinodal at these state points were done from a
quadratic extrapolation of the isotherms. The data points obtained from REMC simulations
was fitted with a quadratic function ($p_0 + a1(\rho-\rho_0) + a2(\rho-\rho_0)^2$),
where $p_0$ and $\rho_0$ are the spinodal pressure and density values. The data and the
fits are shown in FIG.  \ref{fig:LowT_Spinodal}.

\begin{figure}[t]
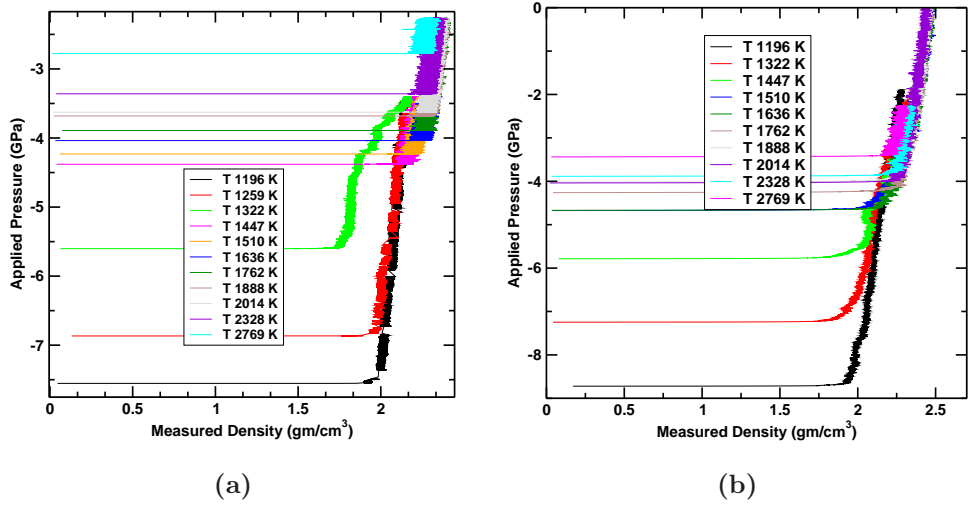

	\begin{tabular}{cc}
		\subfloat[]{\includegraphics[scale=0.51]{images/Si_LLPT/AppliedP_Rho_Rate0.001_Real.eps}}\hspace{5 mm} 
		&\subfloat[]{\includegraphics[scale=0.5]{images/Si_LLPT/AppliedP_Rho_Rate0.100_Real.eps}}\\
	\end{tabular}
	\caption{Applied pressure against measured density for different temperatures from NPT MD
	simulations using the SW potential. The stretching rate in (a) corresponds to $0.1 MPa/ps$
	and in (b) corresponds to $10.0 MPa/ps$.}
	\label{fig:Slow_Fast_Rate}
\end{figure}

As a further check on the spinodal estimate, the tensile limit of the liquid was obtained by
increasing the tensile pressure on the simulation cell at different constant rates. The tensile
limit line is defined as the the locus of maximum tensile stress (negative pressure) a
system can withstand before it fails. At a given temperature the system was first
equilibrated at a high pressure value (for $T < 1510 K$ at $P = -2.26 GPa$ and for $T
> 1510 K$ at $P = 0 GPa$, by performing NPT MD simulation) and then a tensile pressure
was applied which increases at a specified rate. Simulations were performed at four
different constant rates of change of tensile pressure ($0.1 MPa/ps$, $1.0 MPa/ps$, $10.0
MPa/ps$, $50.0 MPa/ps$). When the system reaches its limit of tensile strength, the
system's density decreases drastically towards zero. FIG. \ref{fig:Slow_Fast_Rate} show
the applied pressure against the measured density for a range of temperatures, from which
tensile limit line was obtained. At faster stretching rate ($10 MPa/ps$) the tensile limit
was found to be consistent with the spinodal estimates. For slow stretching rate ($0.1
MPa/ps$) it was found that the system cavitates at higher (less negative) pressure values.
At the intermediate rate, ($1.0 MPa/ps$) the estimated tensile limit line lies between the
estimates obtained from the faster ($10 MPa/ps$) and slow ($0.01 MPa/ps$) stretching
rates. At very high rate of change of tensile pressure ($50.0 MPa/ps$), the tensile limit
was found to be below the spinodal estimates, indicating perhaps that the stretching rates
are faster than microscopic relaxation time scales of the system.

\begin{figure}[b]
	\begin{center}
		\includegraphics[scale=0.55]{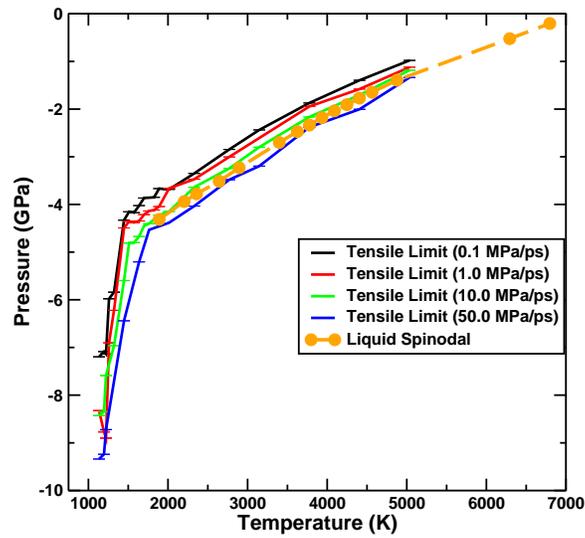}
		\caption{Pressure against temperature showing tensile limits obtained from different
		stretching rates along with the estimated spinodal line from MD simulations using
		the SW potential.}
		\label{Tensile_Limit_Compare}
	\end{center}
\end{figure}

In the FIG. \ref{Tensile_Limit_Compare}, the tensile limit obtained from different
stretching rate along with the spinodal estimate is shown in the $P-T$ plane. At around
$P=-4.0 GPa$, the slope of the tensile limit changes to a bigger value, and based on the
location of the compressibility maximum line, the states near the tensile limit change
from HDL to LDL-like. In the phase diagram, the region of the slope change of the tensile
limit correspond to the region where the TMD line meets the TMinD line and the TMC line
meets the TMinC line. The connection between the change in slope of tensile limit to the
changes in the nature of the states (HDL to LDL-like) is very interesting and needs to be
investigated further.

\begin{figure}[t]
		\includegraphics[scale=0.22]{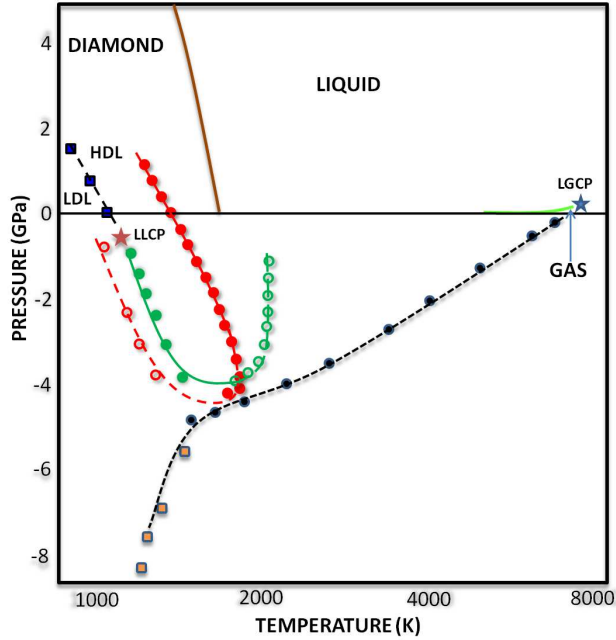}
		\caption{The phase diagram of supercooled silicon in pressure-temperature (P, T)
		plane obtained from simulations using the SW potential. The phase diagram shows the
		the location of	(i) the liquid-crystal phase boundary \cite{Voronin_PRB_2003} -
		brown line, (ii) the liquid-gas phase boundary and critical point - green line and
		blue	star, (iii) the liquid-liquid phase boundary and critical point - blue
		square and brown star, (iv) the liquid spinodal - black circles (v) the tensile
		limit - brown square (vi) the density maximum (TMD) and minimum (TMinD) lines - red
		open and filled circles, and (vii) the compressibility maximum (TMC) and minimum
		(TMinC) line - green closed and open circles. Lines joining TMD and TMinD
		(dot-dashed), TMC and TMinC (solid), Spinodal (black dotted line) are guides to the
		eye.}
		\label{fig:Phase_Diagram}
\end{figure}

\begin{figure}[t]
	\begin{center}
		\includegraphics[scale=0.5]{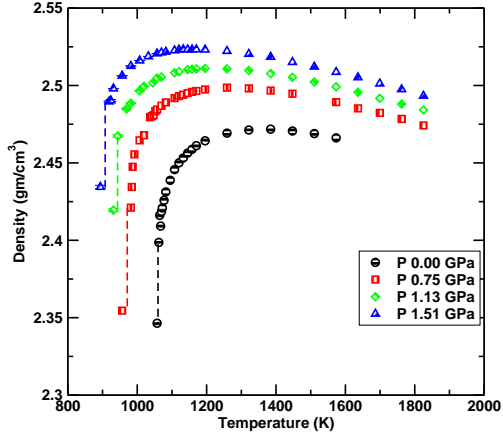}
		\caption{Density against temperature for four different isobars from NPT MD
		simulations using the SW potential. The jumps in the isobars were used to identify
		the liquid-liquid transition line.}
		\label{fig:Coexistence}
	\end{center}
\end{figure}

The complete phase behaviour of supercooled silicon modeled by SW potential is summarised
in FIG. \ref{fig:Phase_Diagram}. The phase diagram includes the liquid-liquid critical point,
liquid-liquid coexistence line (obtained from identifying the jumps in density for small
changes in temperature isobars generated using NPT MD simulations {\it
see} FIG.\ref{fig:Coexistence}), crystal-liquid coexistence line, liquid-gas coexistence
line, the loci of TMD, TMinD, TMC and TMinC along the liquid-spinodal and tensile limit
line. The estimated spinodal is monotonic in pressure {\it vs.} temperature $T$, {\it ie}
not ``reentrant'' as predicted to be the case \cite{Speedy_JPC_1982} if it intersects the
TMD. The TMD, changes slope upon intersection with the TMinC, as analysed in
\cite{sastry_pre_1996}. Evaluating the relevant equation of state data as the TMD
approaches the spinodal is particularly
challenging. From available data, the authors conclude that the TMinC appears to be
smoothly joining with the TMC (line of compressibility maxima) that emanates from the
liquid-liquid critical point. Using the PT MC simulations below the critical temperature
and pressure a line of density minima was also recognised (which was recently observed in
the case of water in experiments and computer simulations \cite{poole_JPCM_2005,
liu_PNAS_2007}) which appears to smoothly join the TMD line, as required by thermodynamic
consistency.


\subsubsection{\bf{Structural and dynamical properties}}

Various structural and dynamical properties of supercooled liquid silicon calculated as
function of pressure and temperature are now summarised. These properties include
relaxation times, structural properties and dynamical properties. In simulations, the
structural relaxation times are calculated from the coherent intermediate scattering
function ($F(q,t)$), which is defined as
\begin{equation}
	F(q,t) = \frac{1}{N}\left < \delta \rho(\vec{q},t) \delta \rho(-\vec{q},0) \right>.
\end{equation}
\noindent
where $\delta \rho(\vec{q},t)$ is the fluctuation in local density (at time $t$) in Fourier
space given by
\begin{equation}
	\rho(\vec{q},t) = \sum_{i=1}^{N} \exp \left(i\ \vec{q} \cdot \vec{r}_{i} (t) \right).
\end{equation}
\noindent The $F(q,t)$ is calculated at a $q$ value corresponding to the first peak of
structure factor S(q). The alpha relaxation time ($\tau_{\alpha}$) is obtained usually as
the time at which the $F(q,t)$ decays by a factor of $e$.

The structural properties are quantified by the S(q) and its Fourier transform, the radial
distribution function $g(r)$. The structure factor is defined as
\begin{equation}
	S(q) = \frac{1}{N} \left <\delta \rho(\vec{q}) \delta \rho(-\vec{q}) \right>.
\end{equation}

The $g(r)$ can be directly calcuated in simulation using the relation
\begin{equation}
	g(r) = \frac{V}{N^{2}} \left \langle \sum_{i} \sum_{j \neq i} \delta (\vec{r} - \vec{r}_{ij} ) \right\rangle.
\end{equation}

In simulation one can also measure the structural order parameters which provides local
structural arrangements. The orientational order among the bonds present in the system is
characterised by local orientational order $Q_l$ which is given by
\begin{equation}
	Q_{l} = \left( \frac{4\pi}{2l+1} \sum_{m=-l}^{m=l} Q^{*}_{lm}(\vec{r}_{i}) Q_{lm}(\vec{r}_{i})\right)^{1/2}.
\end{equation}
\noindent
where $Q_{lm}({\bf r}_i) = (1/N_b(i)) \sum_j Y_{lm}({\bf {\hat r}}_{ij})$, $N_b(i)$ is
the number of neighbours of the particle i and $Y_{lm}({\bf {\hat r}}_{ij})$ are the
spherical harmonics evaluated between the neighbours having a position vector $({\bf {\hat
r}}_{ij})$ \footnote{In \cite{sastry_nmat_2003} the prefactor of $1/N_b$, where
  $N_b$ is the number of nearest neighbours, was not included in the
  definition of $Q_{lm}$. Including this factor makes the peak at
  $Q_3=2.3$ more prominent and does not otherwise change any of the
  reported conclusions.}.


The dynamics of supercooled silicon, in addition to the intermediate scattering function,
has been characterised by the self diffusion coefficient. The self diffusion coefficient
or the diffusivity $D$ is obtained in simulations from the mean square displacement using
the Einstein relation
\begin{equation}
	D = \lim_{t \to \infty} \frac{1}{6t}\left < |\vec r(t) - \vec r(0)|^2 \right >.
\end{equation}
or from the velocity autocorrelation function using the Green-Kubo formula 
\begin{equation}
	D = \frac{1}{3} \int_0^{\infty} \left < \vec v(t) . \vec v(0) \right > dt.
\end{equation}


{\bf Dynamics:} 

\begin{figure}[b]
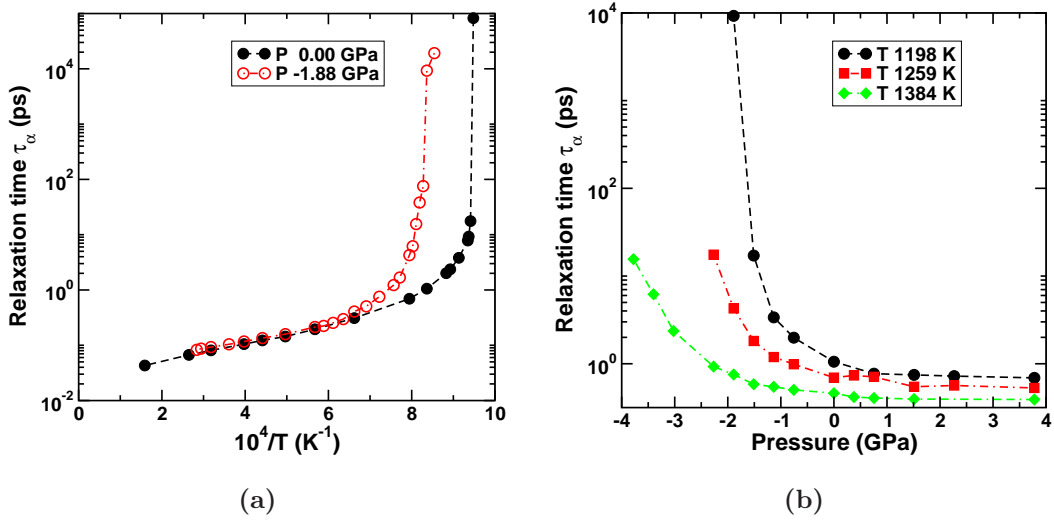

	\begin{tabular}{cc}
		\subfloat[]{\includegraphics[scale=0.5]{images/Properties/Relaxation_Isobars.eps}}\hspace{5 mm}
		&\subfloat[]{\includegraphics[scale=0.51]{images/Properties/Relaxation_Isotherms.eps}}\\
	\end{tabular}
	\caption{(a) Relaxation time ($\tau_{\alpha}$) against inverse temperature at $P = 0 GPa$
	and $P = -1.88GPa$ from NPT MD simulations using the SW potential. (b) Relaxation time
	against pressure at $T = 1198K$, $T = 1259K$ and $T = 1384 K$ from NPT MD simulations using
	the SW potential.}
	\label{fig:Relaxation}
\end{figure}

One of the biggest challenges in the study of supercooled silicon is that, with deep
undercooling the system, not only becomes prone to rapid crystallisation but also the
relaxation times increase very rapidly. Vasisht {\it et al.} studied the relaxation times
at different parts of the phase diagram and found that approaching the liquid-liquid
transition line (below the critical temperature) and the compressibility maxima line
(above the critical temperature), the relaxation time increases in a non-Arrhenius manner.
In the FIG. \ref{fig:Relaxation} (a), the relaxation time as a function of temperature at
two different pressure values (above - $P = 0 GPa$ and below - $P = -1.88GPa$ critical
point) is shown. FIG. \ref{fig:Relaxation} (b) shows relaxation times as a function of
pressure for three different temperatures values.

\begin{figure}[t]
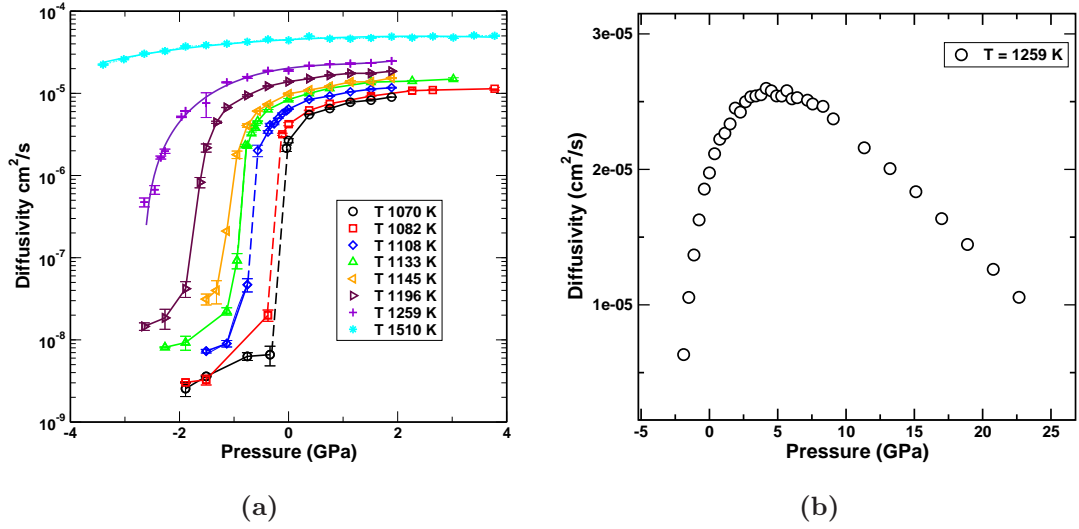

	\begin{tabular}{cc}
		\subfloat[]{\includegraphics[scale=0.5]{images/Properties/Diffusivity_Pressure_Real.eps}}\hspace{5 mm}
		&\subfloat[]{\includegraphics[scale=0.525]{images/Properties/DMax_T0.050.eps}}\\
	\end{tabular}
	\caption{Diffusivity against pressure from NPT MD simulations using the SW potential: (a) For different
	isotherms. Diffusivity decreases with decrease in pressure. (b) For $T =  1259 K$.
	Diffusivity goes through a maximum at around $4.5 GPa$.}
	\label{fig:Diffusivity}
\end{figure}

Vasisht {\it et al.} \cite{Vasisht_nphys_2011} report the
calculation the diffusivity at temperatures below and above the critical temperature. As
the liquid transforms from HDL to LDL the diffusivity is found to change by two orders of
magnitude. The diffusivity is found decrease with decrease in pressure, which is an
anomalous behaviour similar that of water and silica
\cite{errington_nat_2001,shell_PRE_2002}. At higher pressure values diffusivity goes
through a maximum and a return to normal behaviour (FIG. \ref{fig:Diffusivity}). This has
been also observed in {\it ab inito} simulations\cite{Morishita_PRE_2005}.



{\bf Structural properties: }

\begin{figure}[b]
	\begin{tabular}{cc}
		\subfloat[]{\includegraphics[width=0.41\textwidth]{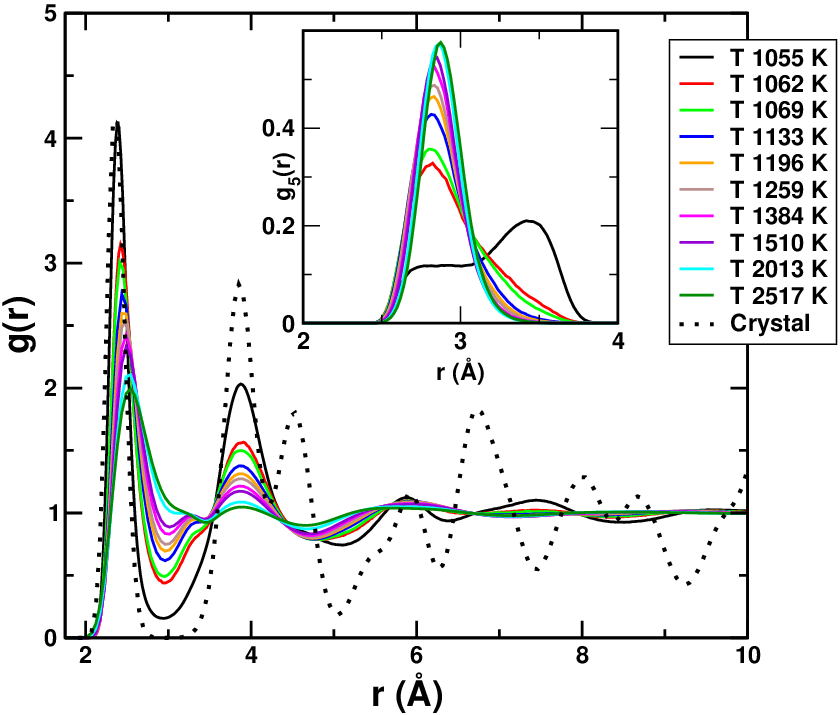}}\hspace{5 mm}
		&\subfloat[]{\includegraphics[width=0.37\textwidth]{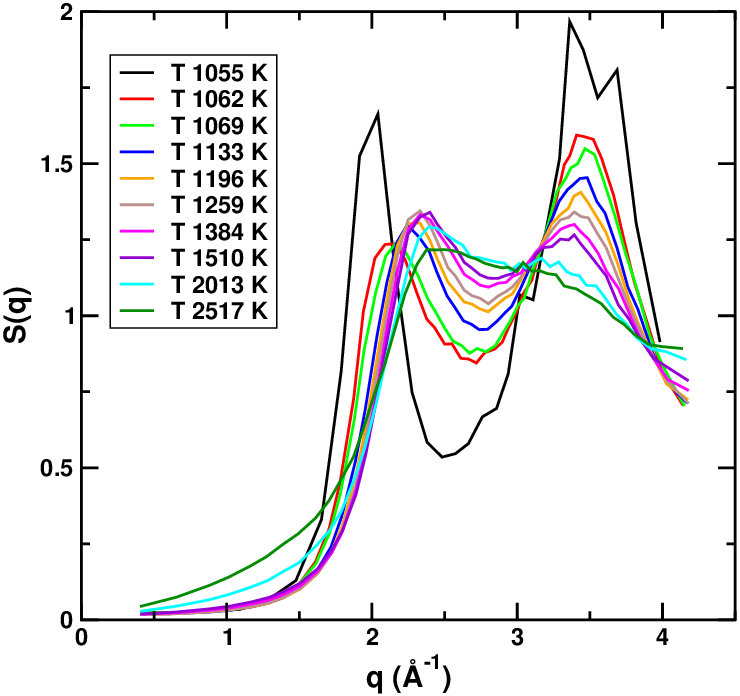}}\\
	\end{tabular}
	\caption{(a) The pair correlation function g(r) and (b) the structure factor S(q) for different temperatures at $P = 0 GPa$ from NPT MD
	simulations using the SW potential. The inset in (a) shows the fifth neighbour distribution.}
	\label{gofr_sofq_P0.00}
\end{figure}

The work of Sastry and Angell \cite{sastry_nmat_2003} showed that the liquid-liquid
transition marks change in structural and dynamic properties. Structurally the high
temperature, HDL is less tetrahedral and has average coordination
number values around $5$ and the low temperature, LDL is more
tetrahedral with average coordination number values around $4$. Vasisht {\it et al.}  have
made a detailed analysis of the $g(r)$ and $S(q)$ which will be summarised in here.  The
$g(r)$ and $S(q)$ along the $P = 0GPa$ and $P = -1.88 GPa$ isobars are shown in the FIG.
\ref{gofr_sofq_P0.00} and FIG. \ref{gofr_sofq_P-0.05} respectively. As the system
transforms from HDL to LDL, the amplitude of the first peak of $g(r)$ (at $r$ $\sim$
$2.6\AA$) was found to increase and the peak shift towards lower values of $r$ approaching
the crystalline peak. Also the amplitude of the first minimum of $g(r)$ decreases indicating a
change towards a first coordination shell similar to that of the crystal. An intermediate
peak or bump due to the presence of the fifth neighbour in the first coordination shell
vanishes with the lowering of temperature. Even though the first coordination shell of
LDL is similar that of the crystal, the higher coordination shells are distinctly
liquid-like, and depart significantly from the crystal $g(r)$. The fifth neighbour
distribution for these state points are shown in the inset of FIG.  \ref{gofr_sofq_P0.00}
and FIG. \ref{gofr_sofq_P-0.05}. The authors find that the $P = 0 GPa$ isobar is above the
critical pressure and hence one sees a discontinuous change of $g(r)$, whereas the $P =
-1.88 GPa$ isobar is below the critical pressure and hence a continuous evolution of structural
change is seen.

\begin{figure}[b]
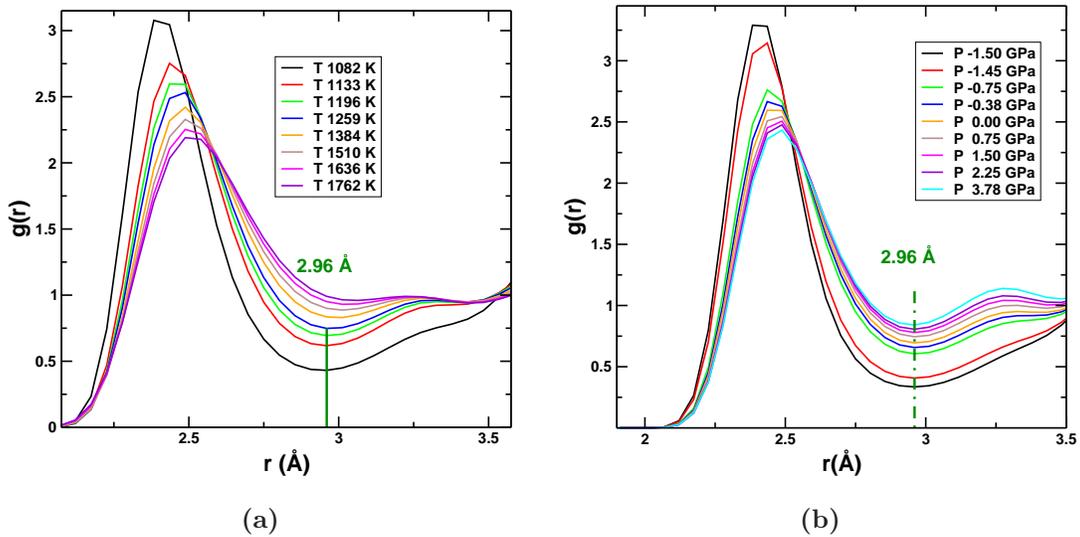

	\begin{tabular}{cc}
		\subfloat[]{\includegraphics[scale=0.5]{images/Properties/gofr_compare_diffT_p0.0_Min.eps}}\hspace{5 mm}
		&\subfloat[]{\includegraphics[scale=0.495]{images/Properties/gofr_compare_diffP_T0.0475.eps}}\\
	\end{tabular}
	\caption{The pair correlation function g(r) from NPT MD simulations using the SW potential: (a) At different temperatures
	at $P = 0 GPa$. The first minimum of $g(r)$ remains unchanged till $T < 1259K$. (b) At
	different pressures at $T = 1196 K$. The first minimum of $g(r)$ remains fairly unchanged
	for a wide range of pressure values.}
	\label{Minima_gofr_P}
\end{figure}
\begin{figure}[t]
	\begin{center}
		\includegraphics[scale=0.6]{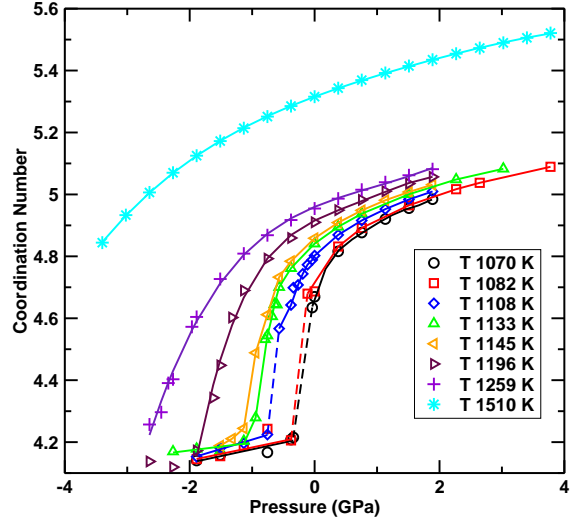}
		\caption{Coordination number against pressure at different temperatures from NPT MD
		simulations using the SW potential. The coordination number for the HDL phase varies
		from $4.6$ to $5.4$. In the LDL phase the coordination number is around 4.2.}
		\label{fig:Cnn_Pres}
	\end{center}
\end{figure}

The coordination number $C_{nn}$ is the number of atoms in the first coordination shell,
and is calculated by integrating the $g(r)$ till its first minimum ($r_{c}$) using the
equation $C_{nn} = \int_{0}^{r_{c}} 4\pi r^{2} \rho g(r) dr$. The coordination number was
found to be very sensitive to the location of the first minimum of $g(r)$. A cutoff of $r_{c}
\sim 2.96\AA$ was found for temperatures less than $1259K$ and for $T>1259K$, the minimum
was found to shift towards higher values of $r$ (FIG. \ref{Minima_gofr_P} (a)). For a
given temperature, with varying pressure, $r_{c}$ was found not to change much (FIG.
\ref{Minima_gofr_P} (b)). In FIG. \ref{fig:Cnn_Pres} the coordination number  is shown as
a function of pressure at different temperatures. For $T<1259K$ the coordination number
was found to vary from $4.6$ to $5.0$ in the HDL phase, which decreases to around $4.2$ in
the LDL phase. At $T = 1510K$, the coordination number varies between $4.8$ and $5.5$ \footnote{
Note that in \cite{sastry_nmat_2003} and \cite{Vasisht_nphys_2011} the integration was
performed up to the first minimum of $\rho 4 \pi r^2 g (r)$, rather than the g (r)
directly, which leads to a small underestimate in the coordination number at high
temperatures and pressures.}.

The coordination number of silicon is a much debated quantity in the literature which we
discuss further later. The large discrepancy between different experimental calculations
of the radial distribution function and of the density leads to large variations in the
calculated coordination number. We discuss this in detail in Sec. VI.

{\bf Relationship between Structure and Dynamics}

\begin{figure}[b]
	\begin{center}
		\includegraphics[scale=0.6]{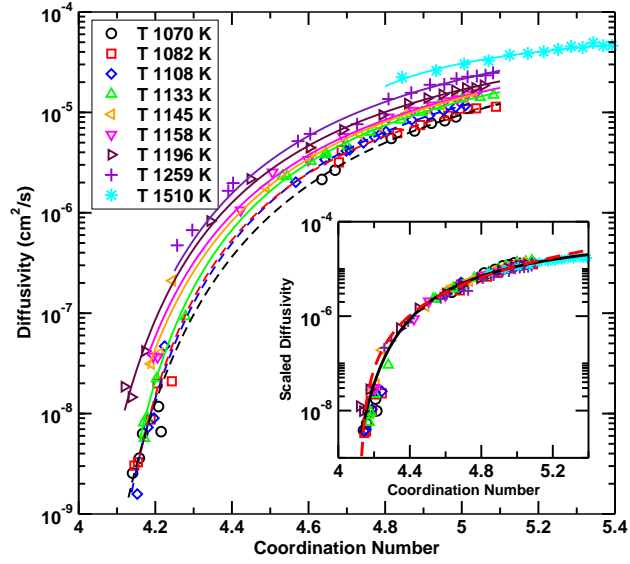}
		\caption{Diffusivity $(D)$ against coordination number $(C_{nn})$ at different
		temperatures from NPT MD simulations using the SW potential. Lines through the data
		points are guides to the eye, and highlight the remarkably similar dependence of
		D on $C_{nn}$ at all temperatures, including those below the critical temperature,
		where both D and $C_{nn}$ change discontinuously. $(Inset)$ Diffusivity (scaled to
		match at $C_{nn} = 4.8$) versus $C_{nn}$, showing data collapse. The solid line is a
		Vogel-Fulcher-Tammann $(VFT)$ fit, with a $C_{nn}$ of vanishing diffusivity $=
		3.86$. The dashed line is a power law fit, with a coordination number of vanishing
		diffusivity $= 4.06$.}
		\label{fig:Diff_Coord}
	\end{center}
\end{figure}

From the data reported above of the coordination number and the diffusivity, it is evident
at a qualitative level that the diffusivity in silicon is correlated with coordination
number, with higher coordination number corresponding to larger diffusivity. It has been
shown in \cite{Vasisht_nphys_2011} that the diffusivity depends quite strongly on
coordination number and has only a weak temperature dependence ({\it see} FIG. \ref{fig:Diff_Coord}). 
Scaling the diffusivity to its value at a fixed $C_{nn}$ in the HDL
phase for different temperatures, a remarkable data collapse is obtained that spans two
distinct phases, a wide range of temperature and pressure, and four decades of change in
diffusivity ({\it see inset} of \ref{fig:Diff_Coord}). The resulting master curve was found to
fit well to a {\it Vogel-Fulcher-Tammann} ($VFT$) form, $D(C_{nn})$ $=$ $D_0$
$exp(-\frac{A}{C_{nn} - n_0})$ with $n_0 = 3.86$, and also to a power law $D(n) = D_0 (n -
n_0)^3$, with $n_0 = 4.06$. These results suggest that the mobility of atoms is strongly tied
to the presence of coordination larger than four, and that regions of higher coordination
number act as ``defects'' that promote faster rearrangements of atomic positions. This
observation is consistent with previous analysis of the role of bifurcated bonds or the
fifth neighbour in determining molecular mobility in water \cite{sciortino_nat_1991,
sciortino_JCP_1992}, though seen here for a remarkably wide temperature and pressure
range.

\section{Electronic Structure} 
\begin{figure}[t]
	\begin{center}
		\includegraphics[scale=0.4]{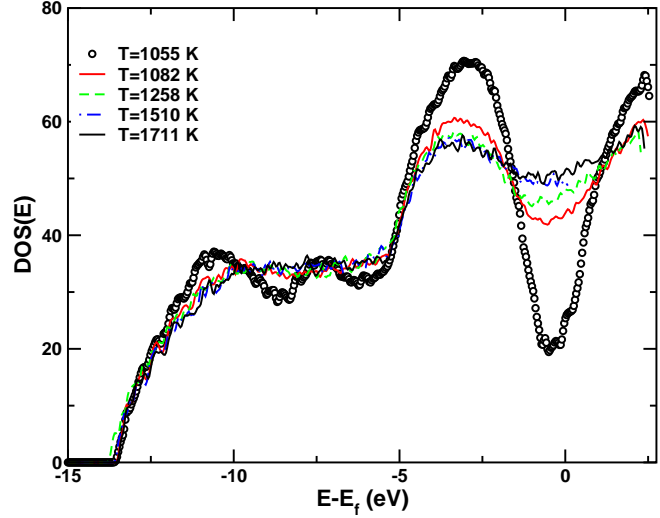}
		\caption{Electronic DOS of the LDL at $1055 K$, HDL at $1082 K$ and high T
		liquid phases from DFT calculations on the MD trajectory obtained using the SW
		potential. [From Ashwin {\it et al.} \cite{ashwin_prl_2004} with permission]}
		\label{Ashwin_EDOS1}
	\end{center}
\end{figure}

\begin{figure}[b]
	\begin{tabular}{cc}
		\subfloat[]{\includegraphics[width=0.51\textwidth]{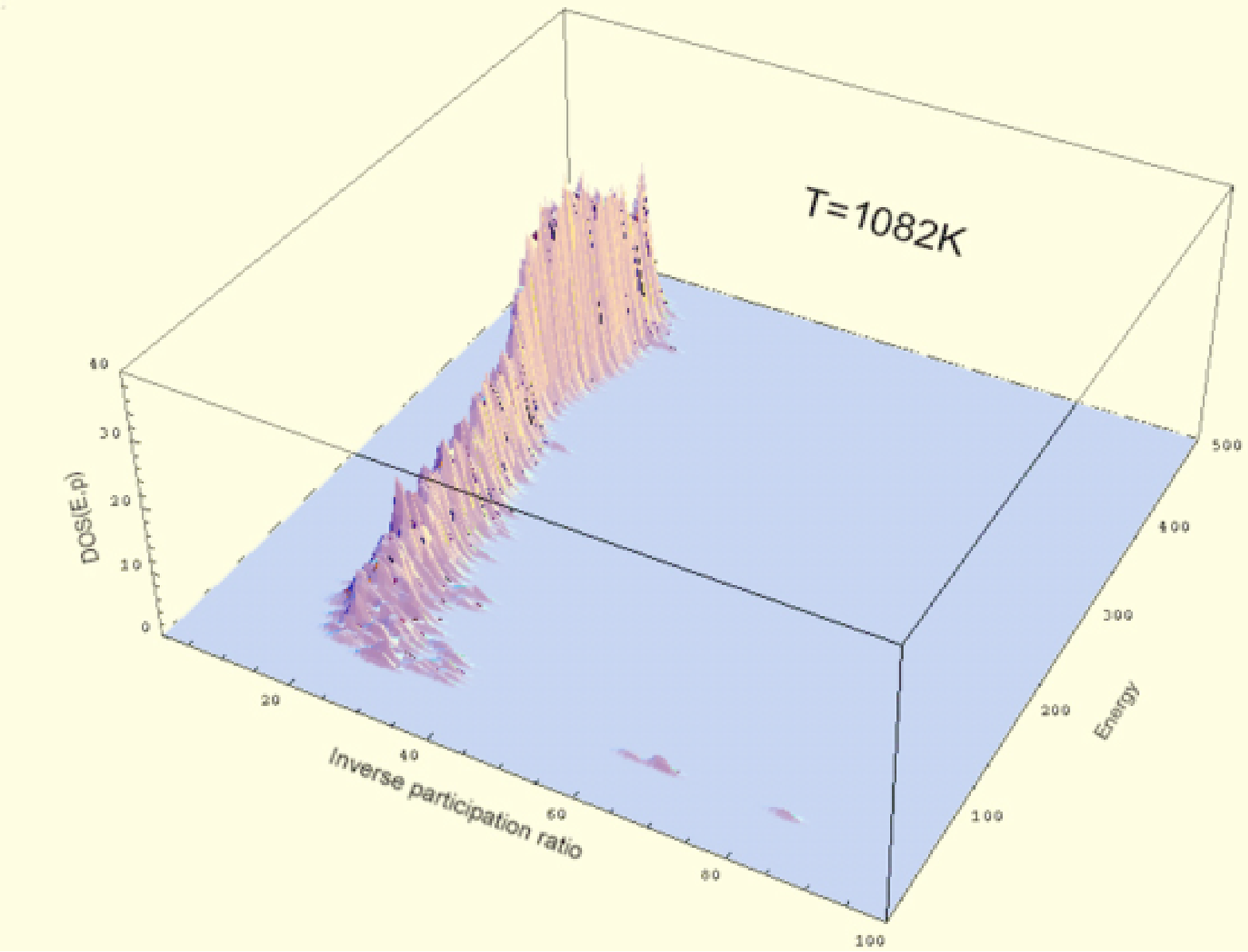}}\hspace{5 mm}
		&\subfloat[]{\includegraphics[width=0.455\textwidth]{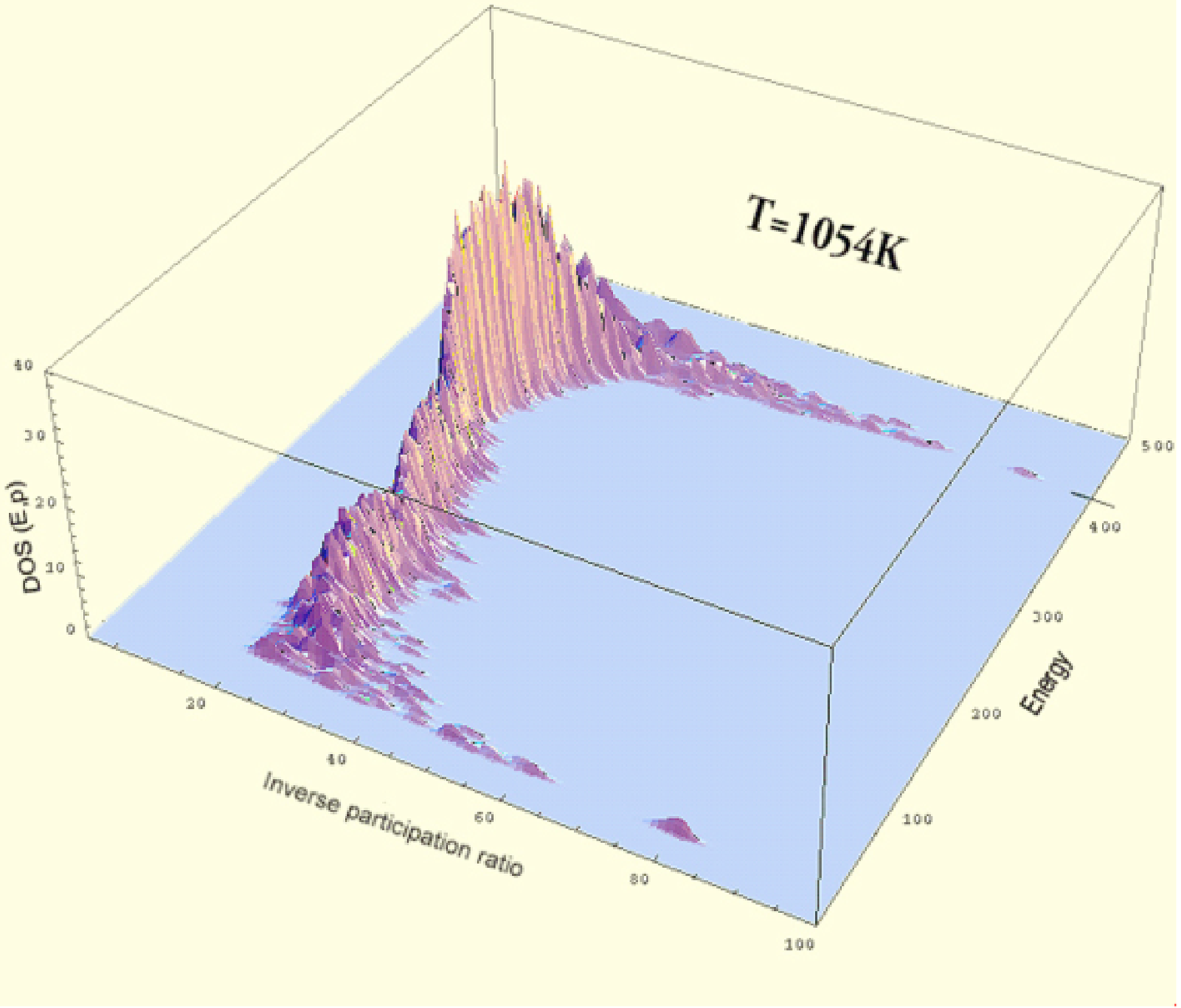}}\\
	\end{tabular}
	\caption{Electronic DOS as a function of energy and inverse participation ration for (a) $T =
	1082 K$ and (b) $T = 1054K$ from DFT calculations on the MD trajectory obtained using the SW
	potential. The states near the Fermi energy at $T = 1054K$ are localised. [From Ashwin {\it
	et al.} \cite{ashwin_prl_2004} and SS Ashwin $PhD$ thesis, JNCASR (2005) with permission.]}
	\label{Ashwin_EDOS2}
\end{figure}

\begin{figure}[t]
	\begin{center}
		\includegraphics[scale=0.4]{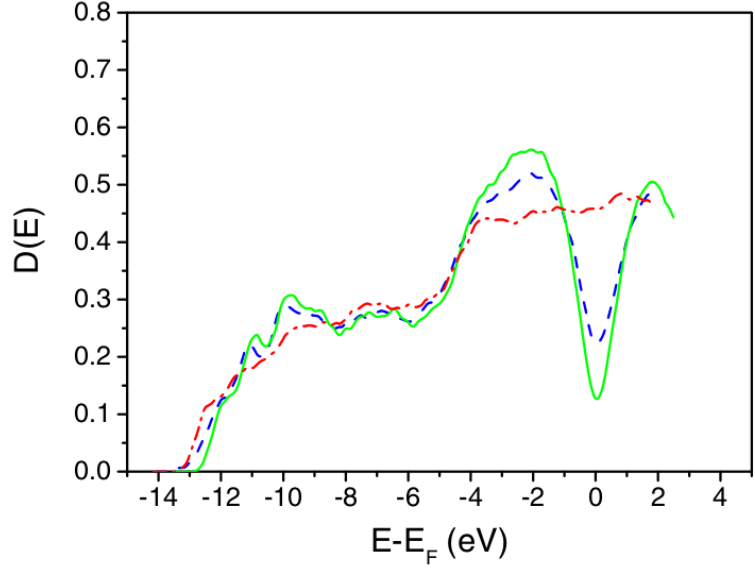}
		\caption{Electronic DOS of the LDL at $1050 K$(green), HDL at $1070 K$(blue) and
		high T liquid at $T_m$ (red) phases from first principles MD (FPMD) simulations.
		[From Jakse {\it et al.} \cite{Jakse_JCP_2008} with permission.]}
		\label{Jakse_EDOS}
	\end{center}
\end{figure}

\begin{figure}[b]
	\begin{center}
		\includegraphics[scale=0.3]{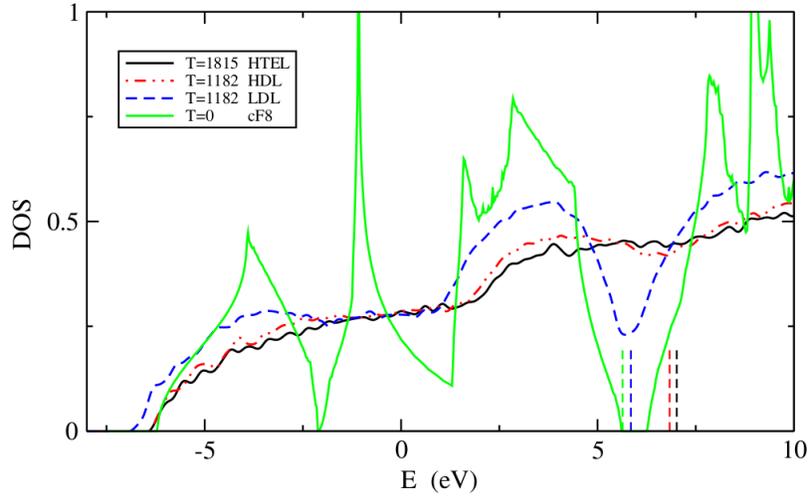}
		\caption{The plot of electronic DOS of the crystal (green), LDL (blue), HDL (red)
		and high T liquid (black) phases from first principles MD (FPMD) simulations. Fermi
		energy $E_F$ for each of the phases is represented by verticle dashed lines. [From
		Ganesh {\it et al.} \cite{Ganesh_PRL_2009} with permission.]}
		\label{Ganesh_EDOS}
	\end{center}
\end{figure}

\begin{figure}[t]
	\begin{center}
		\includegraphics[scale=0.5]{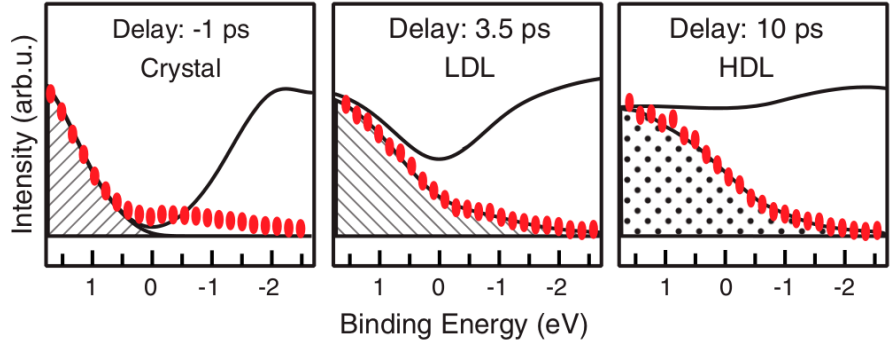}
		\caption{Electronic DOS of the crystal, LDL, and HDL
                  phases. Measured data points for the occupied
                  electronic states are represented by red ovals and
                  black lines are from calculations.[From Beye {\it et
                      al.} \cite{Beye_PNAS_2010} with permission.]}
		\label{Beye_EDOS}
	\end{center}
\end{figure}

Among liquids that may exhibit a liquid-liquid phase transition, a feature that is special
to silicon (though not uniquely so; see earlier discussions) is the change in electronic
properties that accompany the liquid-liquid phase transition. Indeed, this is a feature
that has been exploited in studies from early on in experimental probing of the
transition.  The amorphous-amorphous transition in silicon has been also found to be a
transition from a semiconducting low density state to a metallic high density state. The
liquid form of these phases have shown similar change in the conductivity. Given that the
change in electronic properties has a strong influence on the effective interatomic
interactions, a question has been raised about the extent to which a classical empirical
potential can capture the behaviour seen in silicon. To address a part of this question,
Ashwin {\it et al.} \cite{ashwin_prl_2004}, performed electronic structure calculations,
using an empirical pseudo-potential, for atomic configurations obtained from classical MD
simulations using the SW potential. The electronic density of states $DOS(E)$, obtained
from these calculations is shown in FIG. \ref{Ashwin_EDOS1} and FIG. \ref{Ashwin_EDOS2}. 
Ashwin {\it et al.} found that the DOS remains relatively unchanged at high temperatures
till $T = 1258 K$. A small dip in the DOS(E) at Fermi energy ($E_f$) was found at $T =
1082 K$, near the estimated liquid-liquid transition temperature ($T = 1060K$). In the LDL phase, ($T =
1055K$), even though the DOS(E) remain finite, the authors found a dramatic lowering of
DOS(E) at the Fermi level $E_f$, indicating a change in the conductivity. Further, the
states near the Fermi level become localised in the LDL, as shown in FIG.
\ref{Ashwin_EDOS2} (b), and the conductivity drops by roughly an order of magnitude in
going from HDL to LDL.  Similar features in the DOS(E) have also been found from first
principles calculations \cite{Jakse_JCP_2008, Ganesh_PRL_2009} as shown in FIG.
\ref{Jakse_EDOS} and \ref{Ganesh_EDOS}, indicating that the results are unlikely to be
artifacts arising from the classical nature of the simulations. As described earlier, Beye
{\it et al.} \cite{Beye_PNAS_2010}, have utilised these changes in the electronic DOS to
provide experimental evidence for a liquid-liquid transition. The estimated DOS in their
pump-probe measurements are shown in FIG. \ref{Beye_EDOS}.


\section{Critical Assessment of Classical Simulation Results}

The supercooled phase of silicon has been extensively studied in the past three decades and a
vast amount of knowledge about supercooled silicon comes from computer simulations using
the classical SW model potential, although there have also been many {\it ab initio}
simulations performed in recent years. Any model interaction potential is parameterised so
as to reproduce certain experimentally known properties like phase transition temperature,
crystalline structure {\it etc}. It is difficult for a single empirical interaction
potential to reproduce a wide range of properties in different phases of matter. Unlike
empirical interaction potentials, in {\it ab initio} simulations the effective atomic
interactions are obtained on the fly from quantum mechanical calculations such as Density
Functional Theory ($DFT$) and hence free of parameterisation, although not free of errors
arising from necessary approximations involved in such calculations. But, {\it ab initio}
simulations are computationally very expensive and hence it is difficult to access large
system sizes and time scales using them. These shortcomings becomes a real bottleneck when
one needs to study systems at low temperatures or near a critical point, as is the case
with supercooled silicon. Relaxation times increase with the lowering of temperature and
also with the approach to a critical point. The spatial correlation in the system
increases with the approach to a critical point which makes it necessary to study larger
system sizes (though it must be mentioned that this shortcoming also applies to much of
the classical simulation results presented in this review). In the studies involving supercooled
phases one has to also confront the issue of crystallisation. Since
crystallisation is inherent to a supercooled phase, it is not a feature that can be easily
eliminated without introducing artefacts in the sampling. Hence, it may be necessary
to perform simulations of multiple independent samples to obtain reasonable information on
the metastable states. Given these considerations, in the case of silicon, one must of
necessity perform some of the simulations using a classical empirical potential such as
the SW potential. It is therefore interesting to see to what extent the liquid state
properties evaluated using the SW potential agree with available first principles and
experimental properties.

To this end, in this section we compare various properties obtained in the simulations
using SW potential with the available experimental as well as {\it ab initio} and other model
potential results. These comparisons allow us to understand to what extent the SW silicon
results are applicable to real silicon.  Note that in the comprehensive study of six
different empirical potentials carried out by Balamane {\it et al.}
\cite{balamane_PRB_1992}, the authors looked at various properties at $T=0 K$, but very
little focus was given to the liquid state of silicon.

The Stillinger-Weber potential, by far the most widely used interaction potential for
silicon, comprises of a two- and a three-body interaction potential. The Crystalline phase
of silicon at low pressures is in the diamond cubic structure and it melts into a high
density liquid phase. Stillinger and Weber, after a search through a class of interaction
potentials with two and three body interactions, defined their empirical potential as
follows:

\begin{eqnarray}
u_{SW} = \sum_{i<j}v_2(r_{ij}/\sigma) + 
\sum_{i<j<k} v_3({\bf r}_i/\sigma,{\bf r}_j/\sigma,{\bf r}_k/\sigma), 
\label{eq1}
\end{eqnarray}

\noindent
where $\sigma$ is the diameter of the particles, ${\bf r}_i$ is the position of particle
$i$, and $r_{ij}$ is the distance between particles $i$ and $j$.  The two-body potential
is  short-ranged and has the form

\begin{eqnarray}
v_2(r) &=&
\left \{
\begin{array}{ll}
A \epsilon (B r^{-4}-1)\exp{\left(\frac{1}{r-a}\right)} & r <a  \\
0                                                     & \geq a  
\end{array}
\right .,
\label{eq2}
\end{eqnarray}

\noindent
where $A = 7.049 556 277$, $B = 0.602 224 558 4$, and $a = 1.8$. The repulsive three-body
potential is also short-ranged, and is given by

\begin{eqnarray}
v_3({\bf r}_i,{\bf r}_j,{\bf r}_k) \equiv h(r_{ij}, r_{ik},\theta_{jik}) +  
h(r_{ij}, r_{jk},\theta_{ijk})  \nonumber \\
+  h(r_{ik}, r_{jk},\theta_{ikj}), 
\label{eq3}
\end{eqnarray}

\noindent
where $\theta_{jik}$ is the angle formed by the vectors ${\bf r}_{ij}$
and  ${\bf r}_{ik}$ and

\begin{eqnarray}
h(r_{ij}, r_{ik},\theta_{jik}) &= &\epsilon 
\lambda \exp[\frac{\gamma}{r_{ij}-a} + \frac{\gamma}{r_{ik}-a}] 
\left(\cos\theta_{jik}+\alpha \right)^{2} \nonumber \\
&&\times  H(a-r_{ij}) H(a-r_{ik})  ,
\label{eq4}
\end{eqnarray}

\noindent
where $\lambda =21.0, \gamma = 1.20$, and $H(x)$ is the Heaviside step function. The
choice $\alpha = 1/3$ in $\left(\cos \theta_{jik} + \alpha\right)^{2}$ favors a
tetrahedral arrangement of atoms as found in silicon. The length and energy scales are set
by the choice $\sigma = 2.0951 \AA$, $\epsilon = 50kcal/mol$. The choice of
parameters were identified by taking into account the stable structural arrangement of
the crystal to be cubic diamond, the melting point and the liquid structure. The depth of the
potential $\epsilon$ and diameter of the particle $\sigma$ were chosen such that one
obtains correct lattice spacing and atomisation energy of crystalline Si at 0 K. The
strength of the three body potential is determined by the value of $\lambda$. The two body
part of the potential smoothly goes to zero at the cut off $a$.

\begin{figure}[b]
	\begin{center}
		\includegraphics[scale=0.5]{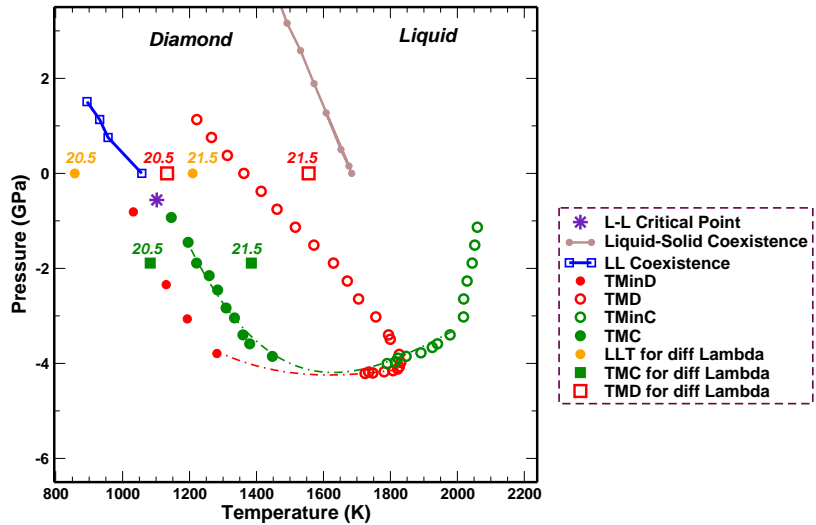}
		\caption{Phase diagram of supercooled silicon (in PT plane) at $\lambda =  20.5, 
		21.0(Si)$ and $21.5$ from MD simulations using the SW potential. The liquid-liquid
		transition points \cite{Molinero_PRL_2006} are shown in orange circles, the density
		maxima point are shown in bold red squares and compressibility maxima points are
		shown in bold opaque green squares. The values of $\lambda$ are stated over the
		symbols.}
		\label{FIG:PD_Changed}
	\end{center}
\end{figure}

\begin{figure}[b]
	\begin{center}
		\includegraphics[scale=0.5]{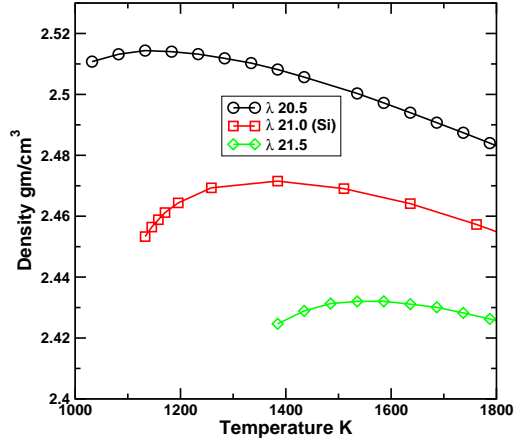}
		\caption{Density against temperature for three different values	of $\lambda$ from
		NPT MD simulations using the SW potential.}
		\label{FIG:Lambda_Rho}
	\end{center}
\end{figure}

The Stillinger-Weber potential is one of the best model potentials for studying the liquid
and supercooled liquid phases of silicon, since the parameters of the model potential are
chosen explicitly to predict the structural properties of real liquid silicon. However,
whether the model faithfully captures temperature variations of thermophysical, structural
and dynamic properties is unclear, and one should expect that the results obtained from
the simulation will be sensitive to the model parameters. The finding of a liquid-liquid
transition in supercooled silicon using the SW potential has been interrogated by Beaucage
and Mousseau \cite{Beaucage_JPCM_2005} based on their study in which they modify the
strength of the three body potential. Beaucage and Mousseau make two observations from
their simulations: (1.) At negative pressures (authors report at $P = -2 GPa$) the
system does not show a first order liquid-liquid transition but instead a second order
transition (2.) For a small change in $\lambda$ the strength of the three body part of
the SW potential (by $5 \%$) the liquid-liquid transition disappears or totally gets
hidden by crystallisation. Hence these authors claim that the nature of the transition in
SW silicon is highly sensitive to model parameters and therefore no strong claims can be
made for what must happen in real silicon based on simulations of the SW potential.  From
the work discussed above, it is clear that there is no phase transition at $P = -2 GPa$,
and examination of well equilibrated data does not reveal any second derivative
singularity to suggest a second order transition as claimed by Beaucage and Mousseau.
Regarding the second point raised by these authors, it is clear that a change in model
parameters will shift the phase diagram around. Importantly, a shift by $5 \%$ in the
value of $\lambda$ changes the liquid-liquid phase transition temperature to higher values
\cite{Molinero_PRL_2006}, to about $T = 1390 K$ at which Beaucage and Mousseau see
crystallisation. It has been observed by many previous studies that crystal nucleation
rates becomes very high when the transition boundary to LDL is crossed
\cite{angell_JNCP_1996, sastry_nmat_2003, Vasisht_nphys_2011}. Hence it is not surprising
that crystallisation is observed at an elevated temperature compared to the normal SW
parameters. In the FIG. \ref{FIG:PD_Changed} we have shown the phase diagram of
supercooled silicon (including the LLT points \cite{Molinero_PRL_2006}, the density maxima
and compressibility maxima at zero pressure) for two different values of $\lambda$ (20.5
and 21.5) along with $\lambda=21.0$. With a small increase in $\lambda$ the phase diagram
shift towards the high temperature (and higher pressure) but the salient features of the
phase diagram do not change. In the FIG. \ref{FIG:Lambda_Rho} we show the change in
density with temperature for different values of $\lambda$. 

\begin{figure}[t]
	\begin{center}
		\includegraphics[scale=0.5]{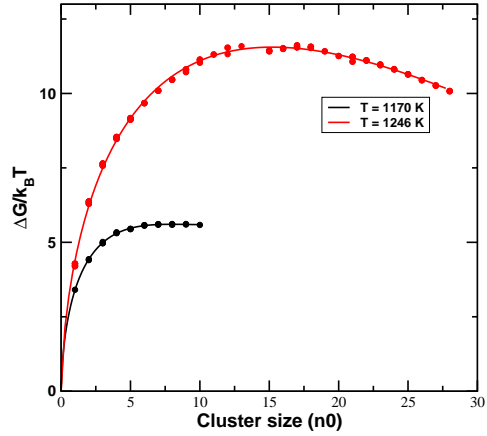}
		\caption{The estimate of Gibbs free energy barrier ($\Delta G/k_BT$) against the 
		largest crystalline cluster size ($n0$)	at $P = 0GPa$ from Umbrella Sampling Monte
		Carlo simulations using the SW potential.}
		\label{FIG:Free_Energy}
	\end{center}
\end{figure}

In the FIG.\ref{FIG:Free_Energy} we show the nucleation barrier
$\Delta G/k_BT$ for the crystallisation of the supercooled liquid as a
function of the size of the largest crystalline cluster
\cite{Vasisht_PREP_2012} for two temperatures, one above and another
below the temperature of compressibility maximum ($T = 1127K$) at $P =
-1.88 GPa$. The nucleation barrier decreases dramatically upon
crossing the compressibility maximum temperature, and hence nucleation 
rates should increase drastically. A detailed work
on change in nucleation rates with strengthening or weakening of
$\lambda$ will shed more light on what Beaucage and Mousseau observe,
but clearly, the qualitative features of the liquid-liquid phase
transition remain intact even with a change in parameters.


Therefore, to evaluate the reliability of the SW potential, we must instead examine how
good the agreement is between properties obtained using it and from experiments or first
principles simulations. Such a comparison is made here on the basis of an extensive
literature search of experiments, first principles simulations and other empirical
potential simulations done on liquid silicon.  We compare results for density ($\rho$),
structure factor (S(q)), radial distribution function (g(r)) and diffusivity (D) obtained
from various reports with the SW silicon data.

Before we go in to the details we would like to mention that the criteria for choosing
the simulation works for the comparison is solely based on the reported temperature range
($T = 1100 K$ to $1700 K$).

\begin{figure}[b]
        \begin{center}
                \includegraphics[scale=0.7]{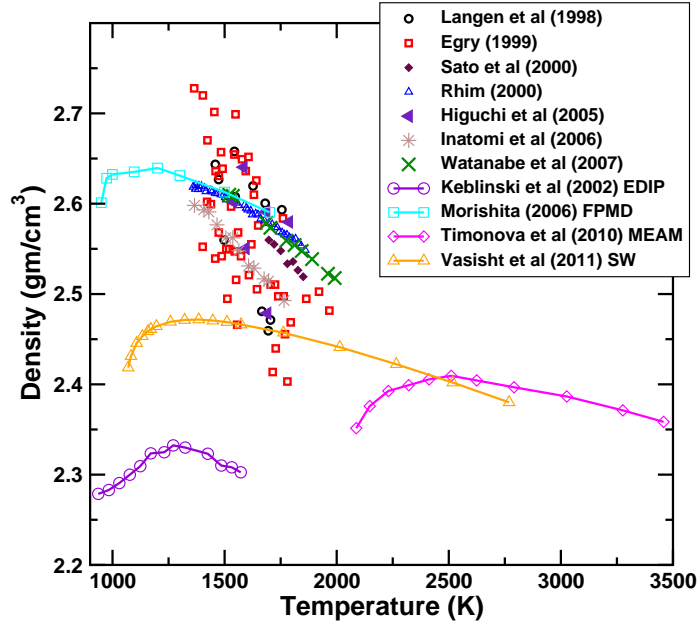}
                \caption{Compilation of density against temperature from different experiments and
		simulations. The experimental data are represented by symbols and the simulation data
		are represented by line and symbol. [From Langen {\it et al.}
		\cite{Langen_JCG_1998}, Egry \cite{Egry_JNCS_1999}, Sato {\it et al.}
		\cite{Sato_IJT_2000}, Rhim \cite{rhim_JCG_2000}, Higuchi {\it et al.}
		\cite{Higuchi_MST_2005}, Inatomi {\it et al.} \cite{Inatomi_IJT_2007} and Watanabe
		{\it et al.} \cite{Watanabe_FarDis_2007}, Keblinski {\it et al.}
		\cite{keblinski_PRB_2002}, Morishita \cite{Morishita_PRL_2006} and Timonova {\it et
		al.} \cite{Timonova_CMS_2010} with permission.]}
		\label{fig:Density_Compare}
        \end{center}
\end{figure}

{\bf Density: } To begin with we compare the density from different experiments and
simulations. We have extracted the $\rho$ from the experimental reports of Langen {\it et
al.} (1998) \cite{Langen_JCG_1998}, Egry (1999) \cite{Egry_JNCS_1999}, Sato
{\it et al.} (2000) \cite{Sato_IJT_2000}, Rhim (2000) \cite{rhim_JCG_2000}, Higuchi {\it
et al.} (2005) \cite{Higuchi_MST_2005}, Inatomi {\it et al.} (2007)
\cite{Inatomi_IJT_2007} and Watanabe {\it et al.} (2007) \cite{Watanabe_FarDis_2007}. We
have used the simulation data reported by Keblinski {\it et al.} (2002)
\cite{keblinski_PRB_2002} using the environment dependent interaction potential, Morishita
(2006) \cite{Morishita_PRL_2006} using {\it ab initio} simulation and Timonova {\it et
al.} (2010) \cite{Timonova_CMS_2010} using the MEAM potential. We show the comparison of
densities from different reports in FIG. \ref{fig:Density_Compare} and we can infer
from this comparison the following points:

\begin{enumerate}
	\item The differences within different experimental values of density are large.
		All the experiments (within  error bar) show monotonically increasing
		density with decrease in temperature till the lowest reported temperature
		\cite{rhim_JCG_2000} of measurement and hence density anomaly is not yet
		observed yet in	experiments.
	\item The SW potential considerably underestimates the densities. However it is
		comparatively better than other estimates from other classical empirical
		potentials \cite{keblinski_PRB_2002, Timonova_CMS_2010}. The density maximum
		as predicted by SW potential at zero pressure is at $T = 1350 K$. The EDIP
		simulations estimate for the density maximum is $T = 1300 K$ and MEAM
		potential estimates at $T = 2500 K$.
	\item Estimates of density from {\it ab initio} simulations compare better with
		the experimentally measured values. The density maximum as predicted by
		FPMD simulation is at $T = 1200 K$ for $P = 0 GPa$.
\end{enumerate}

We next look at the comparison of structural properties.

{\bf Structure factor $S(q)$ and Pair correlation function $g(r)$: } Experiments measure the structure
factor S(q) and the Fourier transform of the S(q) yields the radial distribution function g(r). By
specifying the density as input and integrating the g(r), one obtains the coordination number. From our literature search we find that
agreement between experiments and simulations is as good (or bad) as the agreement between
different experiments. We also find that agreement between experiments and MD simulations
using SW potential is as good as between experiments and first principles simulations. A
detailed comparison of experimental results, first principles results and our results
using SW potential is shown below.

Experimental investigations have measured S(q) down to $1382 K$ \cite{kim_PRL_2005} and not
below. We have chosen a range of temperatures (From T 1100 K to T 1700 K) at normal
pressure. For $T = 1382 K$ we compare the SW simulation data with the experimental data
extracted from Kim {\it et al.} (2005) \cite{kim_PRL_2005}. Between $T = 1447 K$ to
$1667 K$ we compare the SW simulation data with experimental data extracted from Waseda
{\it et al.} (1995) \cite{Waseda_JJAP_1995}, Ansell {\it et al.} (1998)
\cite{Ansell_JPCM_1998}, Kimura {\it et al.} (2001) \cite{Kimura_APL_2001}, Jakse {\it et
al.} (2003) \cite{Jakse_APL_2003}, Higuchi {\it et al.} (2005) \cite{Higuchi_MST_2005},
Kim {\it et al.} (2005) \cite{kim_PRL_2005}, Watanabe {\it et al.} (2007)
\cite{Watanabe_FarDis_2007}, Krishnan {\it et al.} (2007) \cite{krishnan_JNCS_2007}.
Comparison with FPMD and other empirical potential simulations is made from the data
extracted from Jakse {\it et al.} (2003)\cite{Jakse_APL_2003}, Morishita (2006)
\cite{Morishita_PRL_2006}, Wang {\it et al.} (2011) \cite{Wang_PBCM_2011} and
Colakogullari {\it et al.} (2011) \cite{Colakogullari_EPJ_2011}.

\begin{figure}[t]
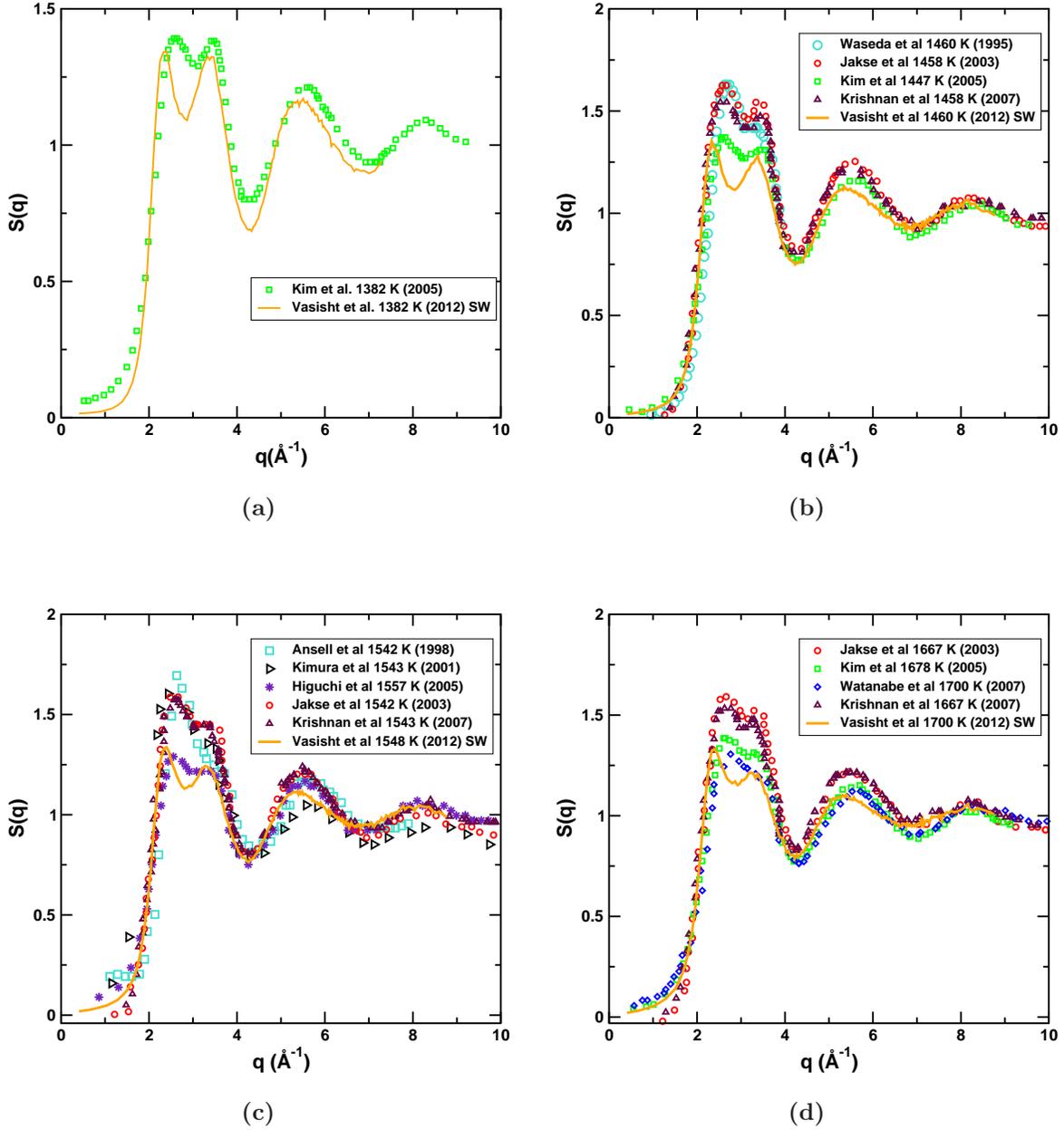

	\begin{tabular}{cc}
		\subfloat[]{\includegraphics[width=0.45\textwidth]{./images/StateofArt/Sofq_T1382_Exp.eps}}\hspace{5 mm}
		& \subfloat[]{\includegraphics[width=0.45\textwidth]{./images/StateofArt/Sofq_T1458_Exp.eps}}\\
		\\
		\subfloat[]{\includegraphics[width=0.45\textwidth]{./images/StateofArt/Sofq_T1550_Exp.eps}}\hspace{5 mm}
		& \subfloat[]{\includegraphics[width=0.45\textwidth]{./images/StateofArt/Sofq_T1700_Exp.eps}}\\
	\end{tabular}
	\caption{Comparison of the structure factor S(q) from NPT MD simulations using the SW potential and from experiments at
	four different temperatures, $T = 1382 K$, $T \approx 1455 K$, $T \approx 1550 K$ and $T
	\approx 1770 K$. [From Waseda {\it et al.} \cite{Waseda_JJAP_1995}, Ansell {\it et al.}
	\cite{Ansell_JPCM_1998}, Kimura {\it et al.} \cite{Kimura_APL_2001}, Jakse {\it et
	al.} \cite{Jakse_APL_2003}, Higuchi {\it et al.} \cite{Higuchi_MST_2005}, Kim {\it et al.}
	\cite{kim_PRL_2005}, Watanabe {\it et al.} \cite{Watanabe_FarDis_2007}, Krishnan {\it et
	al.} \cite{krishnan_JNCS_2007} with permission.]}
	\label{FIG:Sq_Exp}
\end{figure}

We show the comparison of S(q) obtained from SW simulation and experiments in the FIG.
\ref{FIG:Sq_Exp}. and comparison between different simulations is shown in the FIG.
\ref{FIG:Sq_Sim}. We use experimental S(q) of Krishnan {\it et al.} (2007)
\cite{krishnan_JNCS_2007} as a reference for comparing S(q) from different simulations.


We note the following points regarding the structure factor data:

\begin{enumerate}
	\item The differences in structure factor values among different experimental
		reports as well as among different FPMD simulations are significant.

	\item Even though all the experimental data (except Ansell {\it et al.}
		\cite{Ansell_JPCM_1998} at T 1542 K) show similar trends in the structure
		factor, we find a noticeable difference in the amplitudes of the
		structure factor. The trend captured by {\it ab initio} simulations is
		similar to that of SW simulations.

	\item The feature of split peaks in the structure factor is seen in both {\it ab
		initio} and SW simulation. But in comparison with experiments, it seems
		that {\it ab initio} simulation captures better the first peak and the SW
		potential captures better the second peak of structure factor.

	\item The minima and maxima of $S(q)$ for $Q > 4$ $\AA^{-1}$ is well captured by
		the SW potential.
	
	\item The ratio of the first peak to the second peak is better captured by SW potential
		than by the {\it ab initio} simulations.
\end{enumerate}

\par
\begin{figure}[t]
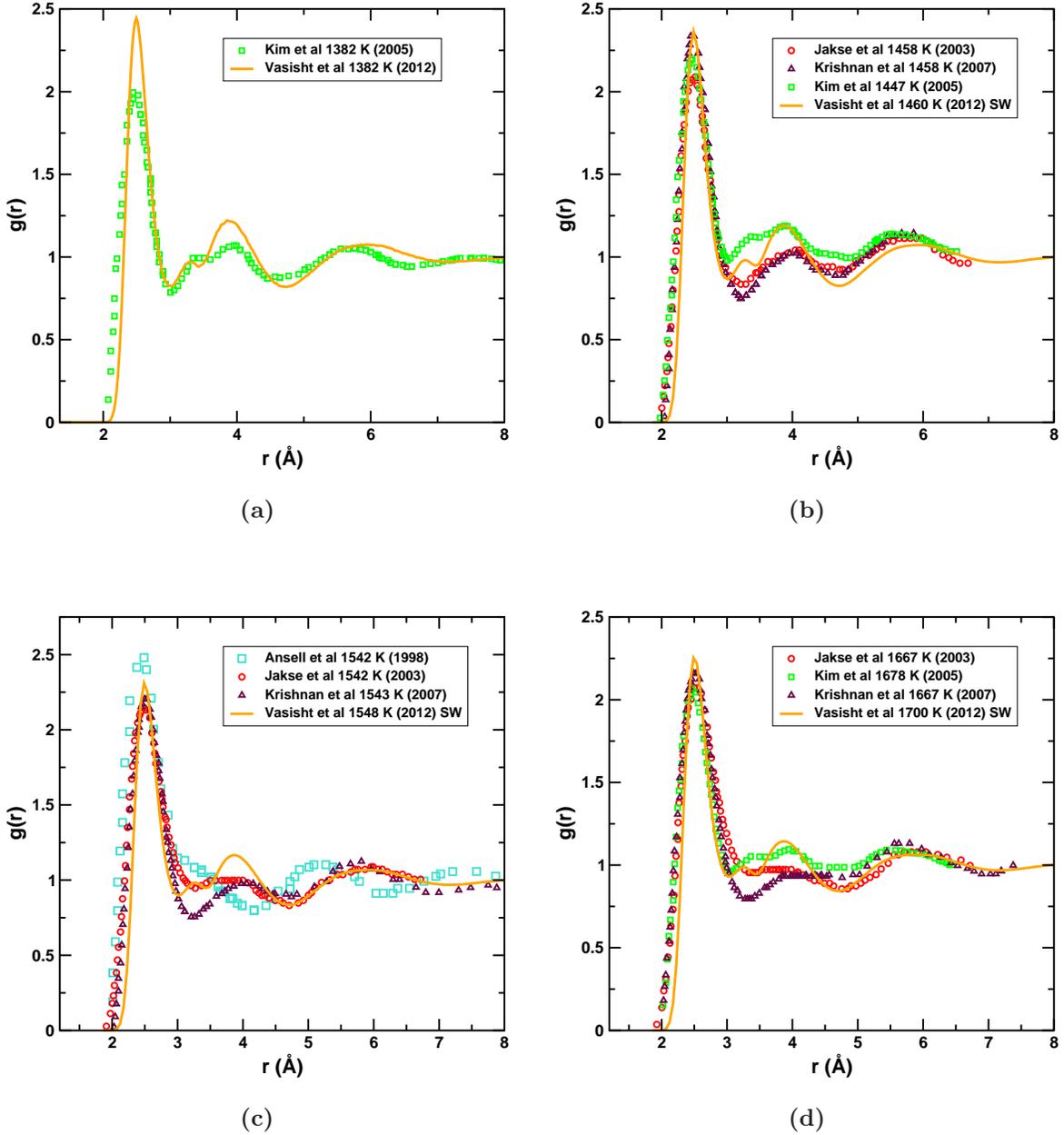


	\begin{tabular}{cc}
		\subfloat[]{\includegraphics[width=0.45\textwidth]{./images/StateofArt/gofr_T1382_Exp.eps}}\hspace{5 mm}
		&\subfloat[]{\includegraphics[width=0.45\textwidth]{./images/StateofArt/gofr_T1458_Exp.eps}}\\
		\\
		\subfloat[]{\includegraphics[width=0.45\textwidth]{./images/StateofArt/gofr_T1542_Exp.eps}}\hspace{5 mm}
		&\subfloat[]{\includegraphics[width=0.45\textwidth]{./images/StateofArt/gofr_T1667_Exp.eps}}\\
	\end{tabular}
	\caption{Comparison of the pair correlation function g(r) from NPT MD simulations using the SW potential and from experiments at
	four different temperatures, $T = 1382 K$, $T \approx 1455 K$, $T \approx 1550 K$ and $T
	\approx 1770 K$. [From Ansell {\it et al.} \cite{Ansell_JPCM_1998}, Jakse {\it et al.}
	\cite{Jakse_APL_2003}, Kim {\it et al.} \cite{kim_PRL_2005}, Krishnan {\it et al.}
	\cite{krishnan_JNCS_2007} with permission.]}
	\label{FIG:gofr_Exp}
\end{figure}

 We now focus on radial distribution function obtained from experiments and
simulations (shown in FIG. \ref{FIG:gofr_Exp} and \ref{FIG:gofr_Sim}) and note the
following points:


\begin{enumerate}
	\item The most noticeable feature is the difference in the first minimum of g(r)
		within different experiments. The coordination number is very sensitive to
		the location of the first minimum of g(r).  The results from Jakse {\it et
		al.} \cite{Jakse_APL_2003} and Krishnan {\it et al.}
		\cite{krishnan_JNCS_2007} are consistent with each other and shows the
		minimum of g(r) at around $3.3 \AA$. The results from Kim
		\cite{kim_PRL_2005} shows the minimum is at around $3 \AA$.
	
	\item Both {\it ab initio} and SW simulations have the first minimum of g(r) at
		around $3 \AA$ (except for Wang {\it et al.} \cite{Wang_PBCM_2011} which
		shows the first minimum of g(r) to be at $3.3 \AA$).
	
	\item The second important difference is in the amplitude of the first peak of
		g(r). Both {\it ab initio} and SW simulations have similar amplitudes of
		the first peak and they are bigger than the observed amplitudes in
		experiments.
	
	\item The intermediate peak between the first and the second prominent peak of g(r) as
		reported by Kim \cite{kim_PRL_2005} is not found in the experiments of
		Jakse \cite{Jakse_APL_2003} and Krishnan \cite{krishnan_JNCS_2007}. But it
		is a prominent feature in both {\it ab initio} and SW simulations.
	
	\item For $r > 3.5 \AA$ (second and higher peaks), the g(r) from SW simulations
		compare better with the	results of Kim {\it et al.} \cite{kim_PRL_2005}
		than the {\it ab initio} results. Results from SW and {\it ab initio}
		simulations are less consistent with the experimental results reported by
		Jakse \cite{Jakse_APL_2003} and Krishnan {\it et al.}
		\cite{krishnan_JNCS_2007}.
\end{enumerate}


A comparison of S(q) and g(r) for the lowest temperature experimentally achieved -$T 1382
K$ \cite{kim_PRL_2005}- ({\it see} FIG. \ref{FIG:Sq_Exp} (a) and \ref{FIG:gofr_Exp} (a)
respectively) shows that even though there are differences in the amplitudes of the
peaks of S(q) and g(r), the SW potential captures all the salient features found in the
experimental results of Kim {\it et al.} \cite{kim_PRL_2005}. 

\begin{figure}[t]
        \begin{center}
                \includegraphics[scale=0.6]{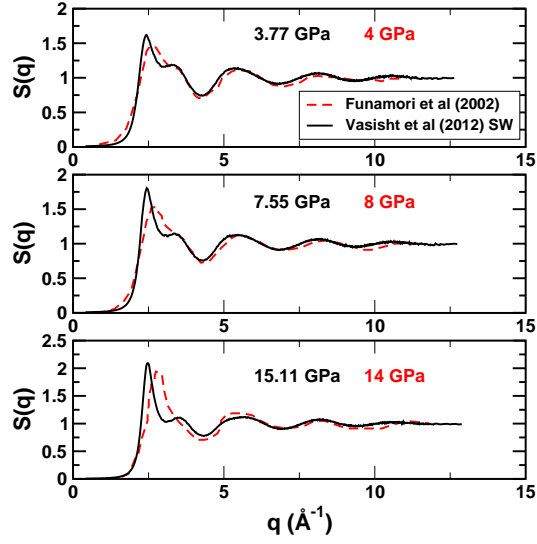}
		\caption{Comparison of the structure factor S(q) from NPT MD simulations using the SW potential with the
		experimental data at high pressure values for $T = 1737 K$ [From Funamori {\it et
		al.} \cite{Funamori_PRL_2002} with permission.]}
                \label{FIG:HighP_Sq}
        \end{center}
\end{figure}

\begin{figure}[b]
        \begin{center}
                \includegraphics[scale=0.6]{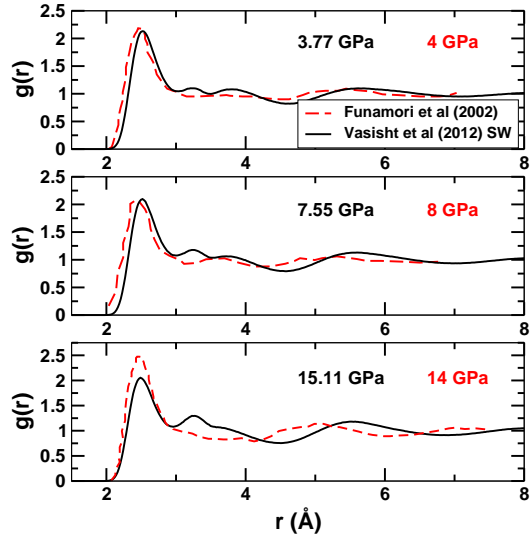}
		\caption{Comparison of the pair correlation function g(r) from NPT MD simulations using the SW potential with the
		experimental data at high pressure values for $T = 1737 K$ [From Funamori {\it et
		al.} \cite{Funamori_PRL_2002} with permission.]}
                \label{FIG:HighP_gr}
        \end{center}
\end{figure}

We also compare the SW data with experimental S(q) and g(r) \cite{Funamori_PRL_2002} at
three different high pressure values at $T = 1737 K$ (FIG. \ref{FIG:HighP_Sq} and FIG.
\ref{FIG:HighP_gr}). We find that at $P = 4 GPa$, the SW data compare reasonably well
with experimental data. But at around $P = 14 GPa$ the SW potential fails to capture the
experimental S(q) and g(r).

From the above comparisons we find that there are noticeable  differences between the
g(r)'s obtained in different experiments as well as from different simulations. We also
find from the comparison of simulation results with experiments, that the SW potential
does as good a job as the {\it ab initio} simulations.

{\bf Coordination number: } The coordination number $C_{nn}$ as discussed before is
calculated by integrating the $g(r)$ till its first minimum ($r_{c}$) using the
equation $C_{nn} = \int_{0}^{r_{c}} 4\pi r^{2} \rho g(r) dr$. Hence the most important
inputs that goes in to the calculation of the $C_{nn}$ is the $r_{c}$ and the density of
the liquid. Hence the differences in the g(r) (both from experiments and simulations) and
the difference in the densities measured and calculated by experiments and simulations,
will reflect in the differences we find in the calculated coordination numbers as it can
seen in the FIG. \ref{fig:Cnn}. To evaluate how much the coordination numbers calculated
by SW potential are affected by the underestimation of the density, we have shown in the FIG.
\ref{fig:Cnn} the coordination numbers calculated using both the densities obtained in
the simulations and the experimental density values (Rhim {\it et al.}
\cite{rhim_JCG_2000}). 
	
\begin{figure}[t]
        \begin{center}
                \includegraphics[scale=0.7]{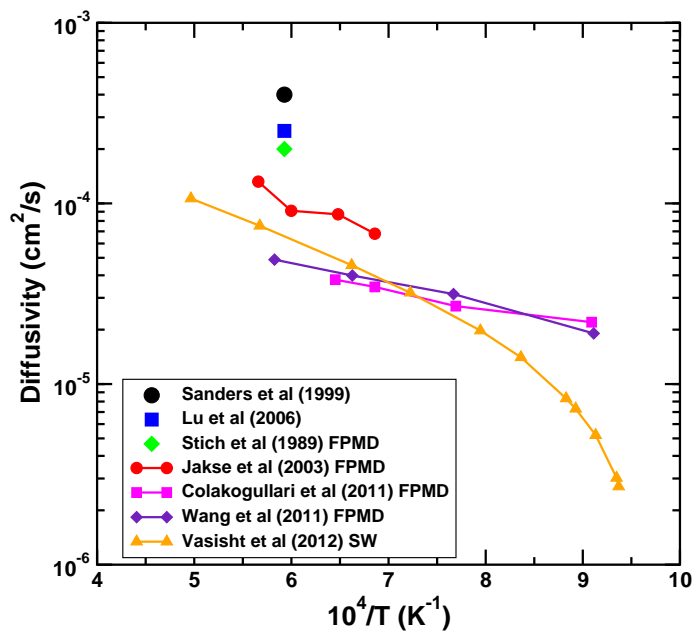}
		\caption{Compilation of Diffusivity against inverse temperature as reported by
		different experimental reports, first principle MD (FPMD) simulations along with the
		simulation results using the SW potential. [From Stich {\it et al.}
		\cite{Stich_1989_PRL}, Jakse {\it et al.} \cite{Jakse_APL_2003},
		Colakogullari {\it et al.} \cite{Colakogullari_EPJ_2011}, Wang {\it et
		al.} \cite{Wang_PBCM_2011}, Sanders {\it et al.} \cite{Sanders_JAP_1999},
		Lu {\it et al.} \cite{Lu_JCG_2006} with permission.]}
                \label{fig:D_Compare}
        \end{center}
\end{figure}

{\bf Diffusivity: } In the FIG. \ref{fig:D_Compare} we compare the temperature dependence
of diffusivity values reported by various simulations and experiments. The FPMD data was
extracted from the reports of Stich {\it el al.} \cite{Stich_1989_PRL}, Jakse {\it et
al.} \cite{Jakse_APL_2003}, Colakogullari {\it et al.} \cite{Colakogullari_EPJ_2011} and
Wang {\it et al.} \cite{Wang_PBCM_2011}. We found experimental report of diffusion
constant only at $T_m$ and we show them in our comparison. We have extracted experimental
values of diffusivity from the works of Sanders {\it et al.} \cite{Sanders_JAP_1999} and
Lu {\it et al.} \cite{Lu_JCG_2006}. We find that only FPMD simulation  of Stich {\it el al.}
\cite{Stich_1989_PRL} predict the diffusivity close to the experimental data.
All other FPMD as well as SW estimates of D (at $T_m$) is smaller than the experimental
value. We find that the SW data is comparable with the FPMD data only down to $T = 1250
K$.

In summary we find that the SW potential, while displaying significant differences with experimental
data for the quantities we have discussed here, does so within deviations that are
comparable to the spread between different experimental results, and these differences are
comparable to those displayed by first principles simulation results. For the different
quantities compared the summary of observations are as follows:

\begin{enumerate}
	\item There are noticeable differences in structure factors and radial
		distribution functions reported by different experimental groups.

	\item The density values reported by different experimental groups have a
		variation of about $10\%$.

	\item The coordination numbers calculated are very sensitive to the location of
		the first minima of g(r) and the density. There is a large spread in the
		coordination numbers reported by different experimental groups, between 5
		and 6.5.

	\item The first principles simulations do not show a better agreement with
		experiments than SW simulations.

	\item The SW potential underestimates the density by $5\%$, which leads to lower
		estimates of the coordination number by $4\%$. Even then these numbers are
		within the error bars of reported experimental	data.

	\item The SW potential estimate of diffusivity (which is comparable to the first
		principle simulations) is approximately a factor of 4 less than the
		experimental value.
\end{enumerate}

It is evident that the most significant shortcoming of the SW potential appears to be in
the estimates of the density. Given the detailed understanding of the SW potential, it may
be interesting to fine tune the potential by varying the strength of the three body
interactions and the range of interactions, which is a useful future direction to pursue.
	
\section{Summary} 

In this review, we have attempted to summarise the course of investigations concerning a
liquid-liquid transition in supercooled silicon, with a significant part of the relevant
results obtained from classical simulations using the Stillinger-Weber potential. We have
attempted to show that such simulations provide a fairly reliable picture of the
behaviour that may be expected for real silicon, through detailed comparison with results
available from experiments and first principles simulations. These results point to the
presence of a liquid-liquid phase transition in silicon, with a liquid-liquid critical
point at negative pressures. The relative inaccessibility of the relevant state points due
to rapid crystallisation poses a challenge both to experimental and simulation
investigations. Nevertheless, it is amply clear that such a transition is of relevance to
understanding processes that are of interest at conditions where the liquid is metastable,
most notably of crystallisation kinetics itself.  Indeed, analysis of crystallisation
kinetics has both historically offered the motivation for the study of such a transition,
as well as a means by which to probe it. As we hope to have illustrated, many clever
experimental approaches have been used and are currently under development to ascertain the
existence of the liquid-liquid transition in silicon. Among the open issues for
simulations to address are the development and validation of more faithful and tractable
model potentials, and a more careful analysis of critical behaviour, such as those that
have recently been performed in the case of water \cite{Liu_JCP_2009,
Kesselring_Arxiv_2011}. A careful analysis of the free energy landscape, and finite size
analysis, along the lines prompted by the work of Limmer and Chandler
\cite{Limmer_JCP_2011} will add valuable insights into the behaviour of metastable liquid
silicon.

\clearpage
\newpage

\clearpage
\newpage

\begin{figure}[b]
	\begin{center}
		\includegraphics[scale=0.5]{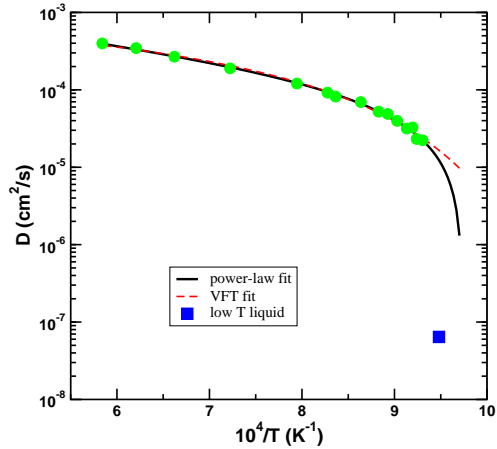}
		\caption{The diffusion coefficient $D$ against the inverse temperature, above and
		below the liquid-liquid transition from MD simulations using the SW potential. In the
		high temperature liquid, the diffusivity show a strongly non-Arrhenius temperature
		dependence. [From Sastry {\it et al.} \cite{sastry_nmat_2003} with permission.]}
		\label{fig:SA_Diffusive}
	\end{center}
\end{figure}

\clearpage
\newpage

\clearpage
\newpage

\begin{figure}[t]
	\begin{tabular}{cc}
		\subfloat[]{\includegraphics[width=0.4\textwidth]{./images/Properties/gofr_diffT_P-0.05.eps}}\hspace{5 mm}
		&\subfloat[]{\includegraphics[width=0.39\textwidth]{./images/Properties/sofq_diffT_P-0.05.eps}}\\
	\end{tabular}
	\caption{(a) The pair correlation function g(r) and (b) the structure factor S(q) for different temperatures at $P = -1.88 GPa$ from NPT MD
	simulations using the SW potential. The inset in (a) shows the fifth neighbour distance 
	distribution.}
	\label{gofr_sofq_P-0.05}
\end{figure}

\clearpage
\newpage

\clearpage
\newpage

\clearpage
\newpage

\begin{figure}[t]
	\begin{tabular}{cc}
		\subfloat[]{\includegraphics[width=0.45\textwidth]{./images/StateofArt/Sofq_T1100.eps}}\hspace{5 mm}
		&\subfloat[]{\includegraphics[width=0.45\textwidth]{./images/StateofArt/Sofq_T1458_Sim.eps}}\\
		\\
		\subfloat[]{\includegraphics[width=0.45\textwidth]{./images/StateofArt/Sofq_T1550_Sim.eps}}\hspace{5 mm}
		&\subfloat[]{\includegraphics[width=0.45\textwidth]{./images/StateofArt/Sofq_T1700_Sim.eps}}\\
	\end{tabular}
	\caption{Comparison of the structure factor S(q) from different simulation works at four different
	temperatures, $T = 1100 K$, $T \approx 1455 K$, $T \approx 1550 K$ and $T \approx 1700 K$.
	We also show the recent experimental S(q) measurements for comparison purposes. [From
	Krishnan {\it et al.} \cite{krishnan_JNCS_2007}, Jakse {\it et al.} \cite{Jakse_APL_2003},
	Morishita \cite{Morishita_PRL_2006}, Wang {\it et al.} \cite{Wang_PBCM_2011} and
	Colakogullari {\it et al.} \cite{Colakogullari_EPJ_2011} with permission.]}
	\label{FIG:Sq_Sim}
\end{figure}

\clearpage
\newpage

\clearpage
\newpage

\begin{figure}[t]
	\begin{tabular}{cc}
		\subfloat[]{\includegraphics[width=0.45\textwidth]{./images/StateofArt/gofr_T1100_Sim.eps}}\hspace{5 mm}
		&\subfloat[]{\includegraphics[width=0.45\textwidth]{./images/StateofArt/gofr_T1458_Sim.eps}}\\
		\\
		\subfloat[]{\includegraphics[width=0.45\textwidth]{./images/StateofArt/gofr_T1542_Sim.eps}}\hspace{5 mm}
		&\subfloat[]{\includegraphics[width=0.45\textwidth]{./images/StateofArt/gofr_T1667_Sim.eps}}\\
	\end{tabular}
	\caption{Comparison of the pair correlation function g(r) from different simulation works at four different temperatures,
	$T = 1100 K$, $T \approx 1455 K$, $T \approx 1550 K$ and $T \approx 1700 K$. We also show
	the recent experimental g(r) measurements for comparison purposes. [From Krishnan {\it et
	al.} \cite{krishnan_JNCS_2007}, Jakse {\it et al.} \cite{Jakse_APL_2003}, Morishita
	\cite{Morishita_PRL_2006}, Wang {\it et al.} \cite{Wang_PBCM_2011} and Colakogullari {\it
	et al.} \cite{Colakogullari_EPJ_2011} with permission.]}
	\label{FIG:gofr_Sim}
\end{figure}

\clearpage
\newpage

\clearpage
\newpage

\bibliography{bibfile}
\end{document}